**Title:**
**Rational Design of Antibiotic Treatment Plans**

Portia M. Mira[1]
Kristina Crona[2]
Devin Greene[2]
Juan C. Meza[1]
Bernd Sturmfels[3]
Miriam Barlow[1]

**Institutional Affiliations:**
[1]School of Natural Science, University of California, Merced
[2] Department of Mathematics and Statistics, American University
[3] Departments of Mathematics, Statistics, and EECS, University of California, Berkeley

**Abstract:**
The development of reliable methods for restoring susceptibility after antibiotic resistance arises has proven elusive. A greater understanding of the relationship between antibiotic administration and the evolution of resistance is key to overcoming this challenge. Here we present a data-driven mathematical approach for developing antibiotic treatment plans that can reverse the evolution of antibiotic resistance determinants. We have generated adaptive landscapes for 16 genotypes of the TEM β-lactamase that vary from the wild type genotype "TEM-1" through all combinations of four amino acid substitutions. We determined the growth rate of each genotype when treated with each of 15 β-lactam antibiotics. By using growth rates as a measure of fitness, we computed the probability of each amino acid substitution in each β-lactam treatment using two different models named the Correlated Probability Model (CPM) and the Equal Probability Model (EPM). We then performed an exhaustive search through the 15 treatments for substitution paths leading from each of the 16 genotypes back to the wild type TEM-1. We identified those treatment paths that returned the highest probabilities of selecting for reversions of amino acid substitutions and returning TEM to the wild type state. For the CPM model, the optimized probabilities ranged between 0.6 and 1.0. For the EPM model, the optimized probabilities ranged between 0.38 and 1.0. For cyclical CPM treatment plans in which the starting and ending genotype was the wild type, the probabilities were between 0.62 and 0.7. Overall this study shows that there is promise for reversing the evolution of resistance through antibiotic treatment plans.

**Blurb:** We explore a novel approach to antibiotic cycling based on evolutionary adaptive landscapes and mathematical optimization of antibiotic treatments. It shows that antibiotic cycling can reverse the evolution of antibiotic resistance.

## Introduction

Antibiotic resistance is an inevitable outcome whenever antibiotics are used. There are many reasons for this: 1) As humans (also as eukaryotes), we are vastly outnumbered by bacteria in nearly all measures, including total population size, biomass, genetic diversity, emigration, and immigration [1]; 2) bacteria can use horizontal gene transfer to share resistance genes across distantly related species of bacteria, including non-pathogens [2]; 3) compared to humans, bacteria have relatively few vulnerable target sites [3]; 4) microbes are the sources of nearly all antibiotics that are used by humans [4]. Given the overwhelming numbers of bacteria, the limited number of target sites, the numerous ways that they can infect humans, and that they have been exposed to naturally occurring antibiotics for billions of years, resistance to antibiotics used by human populations is unavoidable.

Once resistance is present in a bacterial population, it is exceedingly difficult to remove for several reasons. If any amount of antibiotic is present in the environment, antibiotic resistance genes will confer a large fitness advantage [5], and even when antibiotics are not present in an environment, the fitness costs for carrying and expressing resistance genes are small to non-existent [6]. In addition to it being difficult to remove antibiotics from the environment [7], even if humans were to completely abandon the use of antibiotics, resistance would persist for years [8].

Efforts to remove resistance genes from clinical environments by either discontinuing or reducing the use of specific antibiotics for some period of time, either through general reduction of antibiotic consumption or periodic rotations of antibiotics (cycling) have not worked in any reliable or reproducible manner [9]; indeed it would have been surprising if they had worked [10],[11].

Since antibiotic resistance *is* unavoidable, it only makes sense to accept its inevitability and develop methods for mitigating the consequences. One reasonable approach is to rotate the use of antibiotics. This has been implemented in many ways and there are recent studies to model the optimal duration, mixing versus cycling, and how relaxed antibiotic cycles may be and still function as planned [12,13]. However, none of those models have focused on developing a method for designing an optimal succession of antibiotics.

In a previous publication [14], we proposed that susceptibility to antibiotics could be restored by rotating consumption of multiple antibiotics that are a) structurally similar, b) inhibit/kill bacteria through the same target site, and c) result in pleiotropic fitness costs that reduce the overall resistance of bacteria to each other. We presented a proof-of-principle example [14] of how this might work with a series of β-lactam antibiotics in which some of them would select for new amino acid substitutions in the TEM β-lactamase and others that would select reversions in TEM ultimately leading back to the wild type (un-mutated) state. We have focused particularly on β-lactamases because there is often no fitness cost associated with their expression, and they are particularly difficult to remove from clinical microbial populations.

Our current work seeks to identify β-lactam treatment plans that have the highest probability of returning a population expressing a small number of variant TEM genotypes to the wild type state. The wild type TEM-1, and a handful of its descendants, confers resistance to penicillins alone. However, most of the descendants confer resistance to either cephalosporins or penicillins combined with β-lactamase inhibitors (inhibitor resistance), and a few confer resistance to both. Of the 194 clinically identified TEM genotypes that encode unique amino acid sequences [15], 174 (89.7%) differ from the wild type TEM-1 by at most four amino acid substitutions (see Table 1). Our choice of a system that includes four amino acid substitutions is based upon an apparent threshold for amino acid substitutions among functional TEM genotypes. The rarity of the co-existence of cephalosporin resistance and inhibitor resistance and the fact that no single substitution confers both phenotypes suggested that sign epistasis (i.e. reversals of substitutions from beneficial to detrimental) exists as the substitutions that contribute to this dual phenotype are combined.

The ability to apply selective pressures that favor reversions of substitutions within an evolved TEM genotype would increase the number of antibiotics that could be used. To embark upon our effort of determining the best way to do this, we decided to create a model system based upon the TEM-50 genotype, which differs from TEM-1 by four amino acid substitutions. All four substitutions by themselves confer clearly defined resistance advantages in the presence of certain antibiotics. Additionally, TEM-50 is one of the few genotypes that simultaneously confers resistance to cephalosporins and inhibitor combined therapies.

## Results

*From experimental data to mathematical models*

We created all 16 variant genotypes of the four amino acid substitutions found in TEM-50 using site directed mutagenesis (Table 2) and measured the growth rates of 12 replicates of *E.coli* DH5α-E expressing each genotype in the presence of one of fifteen β-lactam antibiotics (Table 3). Each genotype was grown in each antibiotic in 12 replicates. We computed the mean growth rate of those replicates (Table 4) and the variance of each sample, as well as the significance between adjacent genotypes that differed by one amino acid substitution. This was done using one-way ANOVA analysis.

The results are summarized in Figures 1-15, where the arrows in the fitness graphs connect pairs of adjacent genotypes. For each comparison of adjacent genotypes, we indicate the one whose expression resulted in the faster growth by directing the arrowhead towards that genotype, and implying that evolution would proceed in that direction if the two genotypes occurred simultaneously in a population [16,17]. In other words, the node indicated by the arrowhead would increase in frequency and reach fixation in the population, while the other would be lost. Red arrows indicate significance, and black arrows indicate differences that were not statistically significant by ANOVA, but that may still exist if a more sensitive assay was used.

We rank ordered the genotypes (Table 5) in each landscape diagram with a score from 1 to 16, with the genotype promoting the fastest growth receiving a score of “1” and the genotype with the slowest growth a score of “16”. This analysis shows that all genotypes have a score of 5 or better and a score of 13 or worse, in at least one landscape, indicating that there is abundant pleiotropy as antibiotic selective pressures change. That pleiotropy provides a basis for effectively alternating antibiotic to restore the wild type.

Based on the strong patterns of pleiotropy we observed, we reasoned that the choice and the succession of antibiotics were at least as important as other cycling considerations. We formalized our approach to identifying optimal antibiotic treatment paths as follows.

We considered the 15 antibiotics previously mentioned in Table 3: AMP, AM, CEC, CTX, ZOX, CXM, CRO, AMC, CAZ, CTT, SAM, CPR, CPD, TZP, and FEP and their interactions with a bi-allelic 4-locus TEM system $\{0,1\}^4$ where four functionally important amino acid residues involved in the evolution of TEM-50 are considered. The number "1" denotes an amino acid substitution, whereas "0" denotes no substitution at the site. We experimentally determined growth rates for all genotypes in our TEM system at a selected concentration of each antibiotic. Those growth rates depend upon the states of the four amino acid residues. The growth rates for all genotypes in one antibiotic can be represented by a real $2\times2\times2\times2$ tensor $f=(f_{ijkl})$, where $f(a_r)$ is the fitness landscape for the antibiotic $r$. We can identify $f(a_r)$ with a vector whose coordinates are indexed by $\{0,1\}^4$. The resulting 15 vectors, one for each antibiotic, are the rows in Table 4.

Our substitution model $M(f)$ is a function $M:\mathbb{R}^{16}\to\mathbb{R}^{16x16}$ that assigns a transition matrix to each fitness landscape. The rows and columns of $M(f)$ are labeled by the genotypes $\{0,1\}^4$ ordered from 0000 to 1111 according to the degree of lexicographic order. The entries in $M(f(a_r))_{u,v}$ represent the probability that that genotype $u$ is replaced by genotype $v$ under the presence of antibiotic $a_r$. For that reason, the rows of our transition matrices have nonnegative entries and their rows sum to 1.

We require that our transition matrices respect the adjacency structure of the 4-cube, that is, $M(f)_{u,v} = 0$ unless $u$ and $v$ are vectors in $\{0,1\}^4$ that differ in at most one coordinate. In other words, we reasoned that resistant strains are most likely to be in competition with those that express resistance genotypes that are immediately adjacent (vary by a single amino acid substitution).

The combined effect of a sequence $a_1,...,a_k$ of $k$ antibiotics is described by a new transition matrix

$$M(f_{a_1}) \cdot M(f_{a_2}) \cdot \cdots \cdot M(f_{a_k}),$$

obtained as the product of the transition matrices for each drug .
For any genotype u other than 0000, our goal is to find a sequence of antibiotics which maximizes the probability of returning to the wild type. In other words, if we restrict to sequences of length k our goal is to find a sequence of antibiotics $a_1,...,a_k$ which maximizes the matrix entry $M(f(a_1)) \cdot M(f(a_2)) \cdot ... \cdot M(f(a_k))_{u,0000}$.
For each $u$ this requires searching over all $15^k$ antibiotic sequences of length $k$.

*Finding optimal sequences of antibiotics*
We used two substitution models to determine the optimal (most probable) sequences of β-lactams for returning TEM genotypes back to their wild type state. Briefly, the Correlated Probability Model (CPM) allows probabilities to be based upon the actual growth rates. It is given by applying equation (3) to the growth rates in Table 4. The Equal Probability Model (EPM) assumes that beneficial mutations are equally likely and that only the direction of the arrows in Figures 1-15 is important. This means that the matrix entry $M(f)_{u,v}$ is $1/N$ if genotype $u$ has $N$ outgoing arrows and there is an arrow from $u$ to $v$.

A visual summary of the highest probabilities according to the CPM is provided in Figure 16. The CPM provides good estimates if fitness differences between genotypes are small [14,18,19,20]. The EPM has been used in settings where only rank order (as in Table 5) is available [21].

For all sequences of antibiotics of a fixed length (two, three, four, five, and six), we examined the probability that a given genotype is returned to the wild type state. For every starting genotype, we found we were able to return to the wild type genotype with a probability between 0.6 and 1.0 when using the CPM model and a probability of 0.375 and 1.0 when using the EPM model. These results are summarized in Tables 6-9 and Figure 17. These results show the number of paths and their probabilities (Tables 6 and 7) and the substitutions of the most probable paths (Tables 8-11) for returning to the wild type state from various starting points.

Once returned to the wild type state, we identified cycles that would allow for alternation of antibiotics, and allow for some variation through amino acid substitution, but then rapidly return bacteria to the wild type state (Figure 18). Such cycles were possible for path length of two, four, and six and the probabilities of those paths were respectively 0.704, 0.617, 0.617. We found that in the most probable cases, the genotype varied by only one amino acid substitution before reverting back to the wild type state. However, when treatment plans with lower probabilities are considered, we find that more amino acid substitutions in the genotype are allowed.

**Discussion**

In this study, we have developed an experimental approach for measuring pleiotropy and a computational mathematical approach for optimizing antibiotic treatment paths. The experimental approach we developed is rapid and high throughput, and should be applicable to many species of resistant bacteria. The mathematical model we created expresses the problem of antibiotic resistance in general terms, and can therefore be applied to other resistance phenotypes where pleiotropy occurs to identify the antibiotic treatment plans that have the highest probability of reversing the evolution of resistance.

The purpose of this study was to determine whether it is possible to use selective pressures to return TEM-genotypes to the wild type state, as observed in 1963 when TEM-1 was originally isolated. The methods may also be used to select for any particular genotype within our data set. As such, we may select with reasonably high probabilities, resistance genotypes that existed at some prior point in time. To highlight this feature, we have named our software package "Time Machine".

Once given growth rates of adjacent genotypes, Time Machine returned treatment plans that restored the wild type state as observed in 1963 with probabilities greater than 0.6 when using the CPM model and greater than 3/8 (>0.375) when using EPM. These results suggest that when possible it is desirable to use actual growth rates rather than rough ranking data.

The discrete optimization problem motivated by our goal to reverse resistance, or the challenge to build a better time machine, is of independent mathematical interest. Tables 6 and 7 suggest that the maximum probabilities in each row stagnate after a limited number of steps. This is not always the case. We have created an example (see supplemental information) of two substitution matrices on a 3-locus system where the maximum probabilities can be increased indefinitely.

These results show that great potential exists for remediation of antibiotic resistance through antibiotic treatment plans when pleiotropic fitness costs are known for an appropriate set of antibiotics. While developed using a model of Gram-negative antibacterial resistance, this approach could also be used for Gram-positive bacteria and HIV treatment plans.

**Methods**

<u>Experimental methods</u>

*Strains and Cultures*

We expressed 16 mutant constructs of the $bla_{TEM}$ gene in plasmid pBR322 from strain DH5-αE. The 16 genotypes differ at all combinations of four amino acid residues and have been previously described [14]. We grew them overnight (16 hours) in standing cultures and diluted them to a concentration of $1.9X10^5$ as described elsewhere [14].

We transferred 80 *μ*l of each culture to a 384-well plate with one genotype present in each of the 16 rows. The first 12 wells of each row were antibiotic free (controls) and the last 12 wells contained a single antibiotic at an inhibitory, sublethal concentration

After plating, a membrane is placed over the plate and simultaneously incubated/measured in the Eon Microplate Spectrophotometer at a temperature of 25.1°C for 22 hours. This relatively cool (<37º) temperature is used because degradation of the antibiotics is much slower, while the growth rate of the bacteria is still sufficient to capture the complete exponential period of growth over the duration of the experiment. Overall, we have found that a temperature ~25ºC yields more reliable and consistent measurement of growth rates in the presence of antibiotics.

Measurements of cell density (light scattering) at a wavelength of 600 nanometers were automatically collected every 20 minutes after brief agitation to homogenize and oxygenate the culture.

*Growth Rates*

The data obtained from the microplate spectrophotometer is exported to the GrowthRates program to derive the growth rates. In essence, by measuring the optical density at frequent intervals the GrowthRates program can estimate the growth rate,α, through a linear regression algorithm fitting the data from the exponential growth phase. Details can be found in [22] in the section entitled "The Growth Curve" located on pages 233-4. The output of this program for the data we collected was a list $f(a_1), f(a_2), ..., f(a_k)$ of 15 tensors, each of format $2\times2\times2\times2$. These are the rows in Table 4. So if $u \in \{0,1\}^4$ is a genotype, then $f(a_i)_u$ is the fitness of genotype $u$ in the presence of antibiotic $a_i$. This fitness is a growth rate, so we are here using the letter $f$ for a quantity often denoted by α.

One-Way Analysis of Variance (ANOVA) was then used to compare the means of the growth rates obtained, and to determine if there were significant differences between the growth rates of adjacent genotypes.

*Derivation of Correlated Probability Model (CPM)*

Once the growth rates have been determined under various experimental conditions, the next step is to use them to compute fixation probabilities.

If the (multiplicative) absolute fitnesses $W_u$ and $W_v$ of two neighboring genotypes *u* and *v*, differ by a small quantity then the (additive) relative fitness $\ln\left(\frac{W_u}{W_v}\right)$ can be approximated by

1) $$\ln\left(\frac{W_v}{W_u}\right) = T\left(f_v - f_u\right)$$

where $T$ is the generation time. Using a Taylor series approximation,

2) $$\ln\left(\frac{W_v}{W_u}\right) \approx \frac{W_v}{W_u} - 1 .$$

If $W_v > W_u$ , then

3) $$p_{u,v} = \frac{f_v - f_u}{\sum\left(f_{uj} - f_u\right)}$$

is the probability for $v$ to substitute $u$, where $uj$ are the neighbors of $u$ with higher fitness than $u$ [19].

-Derivation of Equal Probability Model (EPM):

According to the EPM model, the probabilities are equal for all beneficial mutations, so that one needs the fitness graphs only for computing the probabilities. The matrix entry $M(f)_{u,v}$ is $1/N$ if genotype $u$ has $N$ outgoing arrows and there is an arrow from $u$ to $v$.

CPM is accurate if fitness differences between genotypes are small, while EPM may provide better estimates if fitness differences are substantial. Indeed, if the fitness effects of all available beneficial mutants exceed some threshold, then fixation probabilities are independent of fitness values [23]. We applied both CPM and EPM, since no complete theory for substitution probabilities exists. Additionally, comparison of two models is useful in learning how sensitive our results are for variation in substitution probabilities.

*Time Machine Programs*

-Optimal antibiotic sequences and pathways of genotypes

Let $M[d]$ denote the $16 \times 16$ transition matrix we derived for the antibiotic labeled $a$. For any sequence $a_1, a_2, ... a_k$ of $k$ antibiotics, we consider the matrix product $M[a_1]M[a_2]...M[a_k]$. This product is also a $16 \times 16$ transition matrix. Its entry in row *u* and column *v* is the fixation probability of genotype $u$ mutating to genotype $v$ under the antibiotic sequence $a_1, a_2, ... a_k$. That probability is a sum of products of entries in the individual matrices $M\left(f\left(a_i\right)\right)$, with one sum for each possible pathway of genotypes from $u$ to $v$. The Time Machine enumerates all $15^k$ antibiotic sequences of length $k$, and it selects all sequences that maximize the entry in row *u* and column *v* of the matrix product. In a subsequent step we then analyze these optimal antibiotic sequences, and for each such sequence, we extract the full list of genotype pathways that contribute.

We implemented this algorithm in the computer algebra software Maple, and we ran it for $k = 2,3,4,5,6$. The running time of the program is slow because of the exponential growth in the number of sequences. At present we do not know whether an efficient algorithm exists for solving our optimization problem for larger values of $k$.

-Cycles of antibiotics

We also used this method to compute cyclical treatment paths in which the starting and ending genotypes were the wild type 0000. The optimization problem we solved was somewhat different from the previous one, in that we focused on obtaining the maximal probabilities of a cycle that includes some substitutions and then returns to the wild type without halting. Halting means that adjacent genotypes in a mutational pathway coincide, which is undesirable.

**Figure Legends**

Figures 1-15

These figures present a visual summary of the adaptive landscape 2x2x2x2 tensors in which each resistance phenotype conferred by each TEM genotype is enumerated. Arrows pointing upward represent addition of a mutation. Arrows pointing downward represent reversions. Red arrows indicate significance between adjacent growth rates as determined by one way ANOVA. Genotypes that confer the most resistance to each antibiotic are shown in red.

Figure 16

Summary of CPM Substitutions with the Highest Probabilities. Each arrow is labeled by the drug or drugs corresponding to the maximal transition probability, taken over all 15 drugs. Each arrow is also labeled by the maximal probability.

From the graph, it is possible to find candidate treatment plans. For example, when starting at genotype 1010 the graph shows that the probability for ending at 0000 is 0.71for the sequence ZOX-TZP (0.71 is the product of the arrow labels). Similarly, when starting at 1111 the probability for ending at 0000 is 0.62 for the sequence CEC-CAZ-TZP-AM. When starting at 0001 the graphs shows that a single drug gives probability at most 0.29, whereas the probability for ending at 0000 for the sequence AMC-CRO-AM (one arrow up, two arrows down) is at least.

This graph can also be used to generate treatment paths that start and end at the same genotype, making possible the development of a fixed treatment plan. For example, from a starting point 0000, the probability for ending at 0000 is 0.62 for the sequence: CEC-CTX-ZOX-CPD-CPR-CAZ-TZP-AM

Figure 17

Summary of Optimal 6 Step CPM and EPM Treatment Paths. Black arrows show transitions present in six step paths computed using both the CPM and the EPM. Red arrows signify transitions found only in optimum paths computed using the CPM whereas blue signify transitions only found using the EPM.

Figure 18

Summary of Optimal CPM 2, 4, and 6 Step Antibiotic Cycles. Two step cycles are shown in red. Four and six step cycles are shown in blue. Four and six step cycles differ only in the number of substitutions and reversions that occur within each cycle. Their probabilities are identical.

**Table 1**

| Number of amino acid substitutions | Number of identified TEM genotypes |
|---|---|
| 1 | 53 |
| 2 | 53 |
| 3 | 37 |
| 4 | 31 |
| 5 | 10 |
| 6 | 2 |
| 7 | 2 |
| 8 | 0 |
| 9 | 0 |
| 10 | 1 |
| 11 | 1 |

**Table 2 Variant Genotypes Created, Binary Codes, Substitutions and (Names of Genotypes Identified in Clinical Isolates)**

| **Number of Substitutions** | **Binary Genotype Code** | **Genotypes with substitutions found in TEM-50** |
|---|---|---|
| 0 | 0000 | No substitutions, (TEM-1) |
| 1 | 1000 | M69L, (TEM-33) |
| 1 | 0100 | E104K, (TEM-17) |
| 1 | 0010 | G238S, (TEM-19) |
| 1 | 0001 | N276D, (TEM-84) |
| 2 | 1100 | M69L, E104K, (Not identified) |
| 2 | 1010 | M69L, G238S, (Not identified) |
| 2 | 1001 | M69L, N276D, (TEM-35) |
| 2 | 0110 | E104K, G238S, (TEM-15) |
| 2 | 0101 | E104K, N276D, (Not identified) |
| 2 | 0011 | G238S, N276D, (Not identified) |
| 3 | 1110 | M69L, E104K, G238S, (Not identified) |
| 3 | 1101 | M69L, E104K, N276D, (Not Identified) |
| 3 | 1011 | M69L, G238S, N276D, (Not identified) |
| 3 | 0111 | E104K, G238S, N276D, (Not identified) |
| 4 | 1111 | M69L, E104K, G238S, N276D, (TEM-50) |

**Table 3 β-lactam Antibiotics used for this study**

| β-lactam Antibiotic | FDA approval | Antibiotic Group |
|---|---|---|
| Ampicillin (AMP) | 1963 | Aminopenicillin |
| Amoxicillin (AM) | 1972 | Aminopenicillin |
| Cefaclor (CEC) | 1979 | Cephalosporin |
| Cefotaxime (CTX) | 1981 | Cephalosporin |
| Ceftizoxime (ZOX) | 1983 | Cephalosporin |
| Cefuroxime (CXM) | 1983 | Cephalosporin |
| Ceftriaxone (CRO) | 1984 | Cephalosporin |
| Amoxicillin + Clavulanic acid (AMC) | 1984 | Penicillin derivative + β-Lactamase inhibitor |
| Ceftazidime (CAZ) | 1985 | Cephalosporin |
| Cefotetan (CTT) | 1985 | Cephalosporin |
| Ampicillin + Sulbactam (SAM) | 1986 | Penicillin derivative + β-Lactamase inhibitor |
| Cefprozil (CPR) | 1991 | Cephalosporin |
| Cefpodoxime (CPD) | 1992 | Cephalosporin |
| Pipercillin + Tazobactam (TZP) | 1993 | Penicillin derivative + β-Lactamase inhibitor |
| Cefepime (FEP) | 1996 | Cephalosporin |

**Table 4 Average Growth Rates ( x $10^{-3}$): the rows are the fitness landscapes**

| | 0000 | 1000 | 0100 | 0010 | 0001 | 1100 | 1010 | 1001 |
|---|---|---|---|---|---|---|---|---|
| **AMP** | 1.851 | 1.570 | 2.024 | 1.948 | 2.082 | 2.186 | 0.051 | 2.165 |
| **AM** | 1.778 | 1.720 | 1.448 | 2.042 | 1.782 | 1.557 | 1.799 | 2.008 |
| **CEC** | 2.258 | 0.234 | 2.396 | 2.151 | 1.996 | 2.150 | 2.242 | 0.172 |
| **CTX** | 0.160 | 0.185 | 1.653 | 1.936 | 0.085 | 0.225 | 1.969 | 0.140 |
| **ZOX** | 0.993 | 1.106 | 1.698 | 2.069 | 0.805 | 1.116 | 1.894 | 1.171 |
| **CXM** | 1.748 | 0.423 | 2.940 | 2.070 | 1.700 | 2.024 | 1.911 | 1.578 |
| **CRO** | 1.092 | 0.830 | 2.880 | 2.554 | 0.287 | 1.407 | 3.173 | 0.540 |
| **AMC** | 1.435 | 1.417 | 1.672 | 1.061 | 1.573 | 1.377 | 1.538 | 1.351 |
| **CAZ** | 2.134 | 0.288 | 2.042 | 2.618 | 2.656 | 2.630 | 1.604 | 0.576 |
| **CTT** | 2.125 | 3.238 | 3.291 | 2.804 | 1.922 | 0.546 | 2.883 | 2.966 |
| **SAM** | 1.879 | 2.198 | 2.456 | 0.133 | 2.533 | 2.504 | 2.308 | 2.570 |
| **CPR** | 1.743 | 1.553 | 2.018 | 1.763 | 1.662 | 0.223 | 0.165 | 0.256 |
| **CPD** | 0.595 | 0.432 | 1.761 | 2.604 | 0.245 | 0.638 | 2.651 | 0.388 |
| **TZP** | 2.679 | 2.709 | 3.038 | 2.427 | 2.906 | 2.453 | 0.172 | 2.500 |
| **FEP** | 2.590 | 2.067 | 2.440 | 2.393 | 2.572 | 2.735 | 2.957 | 2.446 |
| | **0110** | **0101** | **0011** | **1110** | **1101** | **1011** | **0111** | **1111** |
| **AMP** | 2.033 | 2.198 | 2.434 | 0.088 | 2.322 | 0.083 | 0.034 | 2.821 |
| **AM** | 1.184 | 1.544 | 1.752 | 1.768 | 2.247 | 2.005 | 0.063 | 2.047 |
| **CEC** | 2.230 | 1.846 | 2.648 | 2.640 | 0.095 | 0.093 | 0.214 | 0.516 |
| **CTX** | 2.295 | 0.138 | 2.348 | 0.119 | 0.092 | 0.203 | 2.269 | 2.412 |
| **ZOX** | 2.138 | 2.010 | 2.683 | 1.103 | 1.105 | 0.681 | 2.688 | 2.591 |
| **CXM** | 2.918 | 2.173 | 1.938 | 1.591 | 1.678 | 2.754 | 3.272 | 2.923 |
| **CRO** | 2.732 | 0.656 | 3.042 | 2.740 | 0.751 | 1.153 | 0.436 | 3.227 |
| **AMC** | 0.073 | 1.625 | 1.457 | 1.307 | 1.914 | 1.590 | 0.068 | 1.728 |
| **CAZ** | 2.924 | 2.756 | 2.688 | 2.893 | 2.677 | 1.378 | 0.251 | 2.563 |
| **CTT** | 3.082 | 2.888 | 0.588 | 3.193 | 3.181 | 0.890 | 3.508 | 2.543 |
| **SAM** | 0.083 | 2.437 | 0.094 | 2.528 | 3.002 | 2.886 | 0.094 | 3.453 |
| **CPR** | 2.042 | 2.050 | 1.785 | 1.811 | 0.239 | 0.221 | 0.218 | 0.288 |
| **CPD** | 2.910 | 1.471 | 3.043 | 0.963 | 0.986 | 1.103 | 3.096 | 3.268 |
| **TZP** | 2.528 | 3.309 | 0.141 | 0.609 | 2.739 | 0.093 | 0.143 | 0.171 |
| **FEP** | 2.652 | 2.808 | 2.832 | 2.796 | 2.863 | 2.633 | 0.611 | 3.203 |

**Table 5 Rank Order of Genotypes in Each β-Lactam Antibiotic (Derived From Table 4)**

| Antibiotic | 0000 | 1000 | 0100 | 0010 | 0001 | 1100 | 1010 | 1001 | 0110 | 0101 | 0011 | 1110 | 1101 | 1011 | 0111 | 1111 |
|---|---|---|---|---|---|---|---|---|---|---|---|---|---|---|---|---|
| AMP | 11 | 12 | 9 | 10 | 7 | 5 | 15 | 6 | 8 | 4 | 2 | 13 | 3 | 14 | 16 | 1 |
| AM | 8 | 11 | 14 | 3 | 7 | 12 | 6 | 4 | 15 | 13 | 10 | 9 | 1 | 5 | 16 | 2 |
| CEC | 4 | 12 | 3 | 7 | 9 | 8 | 5 | 14 | 6 | 10 | 1 | 2 | 15 | 16 | 13 | 11 |
| CTX | 11 | 10 | 7 | 6 | 16 | 8 | 5 | 12 | 3 | 13 | 2 | 14 | 15 | 9 | 4 | 1 |
| ZOX | 14 | 11 | 8 | 5 | 15 | 10 | 7 | 9 | 4 | 6 | 2 | 3 | 12 | 16 | 1 | 3 |
| CXM | 11 | 16 | 2 | 7 | 12 | 8 | 10 | 15 | 4 | 6 | 9 | 14 | 13 | 5 | 1 | 3 |
| CRO | 10 | 11 | 4 | 7 | 16 | 8 | 2 | 14 | 6 | 13 | 3 | 5 | 12 | 9 | 15 | 1 |
| AMC | 9 | 10 | 3 | 14 | 6 | 11 | 7 | 12 | 15 | 4 | 8 | 13 | 1 | 5 | 16 | 2 |
| CAZ | 10 | 15 | 11 | 8 | 6 | 7 | 12 | 14 | 1 | 3 | 4 | 2 | 5 | 13 | 16 | 9 |
| CTT | 12 | 3 | 2 | 10 | 13 | 16 | 9 | 7 | 6 | 8 | 15 | 4 | 5 | 14 | 1 | 11 |
| SAM | 12 | 11 | 8 | 13 | 5 | 7 | 10 | 4 | 16 | 9 | 14 | 6 | 2 | 3 | 15 | 1 |
| CPR | 7 | 9 | 3 | 6 | 8 | 13 | 16 | 11 | 2 | 1 | 5 | 4 | 12 | 14 | 15 | 10 |
| CPD | 13 | 14 | 7 | 6 | 16 | 12 | 5 | 15 | 4 | 8 | 3 | 11 | 10 | 9 | 2 | 1 |
| TZP | 6 | 5 | 2 | 10 | 3 | 9 | 12 | 8 | 7 | 1 | 15 | 11 | 4 | 16 | 14 | 13 |
| FEP | 10 | 15 | 13 | 14 | 11 | 7 | 2 | 12 | 8 | 5 | 4 | 6 | 3 | 9 | 16 | 1 |
| Best value | 4 | 3 | 2 | 3 | 3 | 5 | 2 | 4 | 1 | 1 | 1 | 2 | 1 | 3 | 1 | 1 |
| Worst value | 14 | 16 | 14 | 14 | 16 | 16 | 15 | 15 | 16 | 13 | 15 | 14 | 15 | 16 | 16 | 13 |
| Median value | 10 | 11 | 7 | 7 | 9 | 8 | 7 | 12 | 6 | 6 | 4 | 6 | 5 | 9 | 15 | 2 |

**Table 6. Maximum Probability and Number of Paths Using CPM**

| Starting Genotype | 1 Step | No. of paths | 2 Step | No. of paths | 3 Step | No. of paths | 4 Step | No. of paths | 5 Step | No. of paths | 6 Step | No. of paths |
|---|---|---|---|---|---|---|---|---|---|---|---|---|
| 1000 | 1.0 | 1 | 1.0 | 3 | 1.0 | 7 | 1.0 | 15 | 1.0 | 31 | 1.0 | 63 |
| 0100 | 0.617 | 1 | 0.617 | 6 | 0.617 | 36 | 0.617 | 219 | 0.617 | 1360 | 0.617 | 8568 |
| 0010 | 0.715 | 1 | 0.715 | 2 | 0.715 | 3 | 0.715 | 4 | 0.715 | 5 | 0.715 | 6 |
| 0001 | 0.287 | 1 | 0.287 | 1 | 0.592 | 2 | 0.592 | 8 | 0.726 | 2 | 0.726 | 4 |
| 1100 | - | | 0.617 | 3 | 0.617 | 18 | 0.617 | 108 | 0.617 | 657 | 0.617 | 4110 |
| 1010 | - | | 0.715 | 1 | 0.715 | 6 | 0.715 | 27 | 0.715 | 112 | 0.715 | 453 |
| 1001 | - | | 0.559 | 1 | 0.559 | 4 | 0.726 | 1 | 0.726 | 2 | 0.729 | 1 |
| 0110 | - | | 0.617 | 1 | 0.617 | 10 | 0.617 | 78 | 0.617 | 555 | 0.617 | 3805 |
| 0101 | - | | 0.592 | 1 | 0.592 | 9 | 0.612 | 1 | 0.612 | 9 | 0.617 | 34 |
| 0011 | - | | 0.361 | 1 | 0.361 | 9 | 0.586 | 2 | 0.600 | 2 | 0.617 | 8 |
| 1110 | - | | - | | 0.617 | 2 | 0.617 | 24 | 0.617 | 215 | 0.617 | 1720 |
| 1101 | - | | - | | 0.592 | 2 | 0.592 | 24 | 0.617 | 12 | 0.617 | 252 |
| 1011 | - | | - | | 0.532 | 1 | 0.532 | 1 | 0.684 | 1 | 0.690 | 1 |
| 0111 | - | | - | | 0.586 | 1 | 0.600 | 1 | 0.617 | 4 | 0.617 | 84 |
| 1111 | - | | - | | - | - | 0.617 | 4 | 0.617 | 72 | 0.617 | 906 |

### Table 7. Maximum Probability and Number of Paths Using EPM

| Starting Genotype | 1 Step | No. of paths | 2 Step | No. of paths | 3 Step | No. of paths | 4 Step | No. of paths | 5 Step | No. of paths | 6 Step | No. of paths |
|---|---|---|---|---|---|---|---|---|---|---|---|---|
| 1000 | 1.0 | 1 | 1.0 | 3 | 1.0 | 7 | 1.0 | 15 | 1.0 | 31 | 1.0 | 63 |
| 0100 | 0.33 | 1 | 0.33 | 6 | 0.33 | 39 | 0.38 | 1 | 0.46 | 1 | 0.46 | 9 |
| 0010 | 0.50 | 1 | 0.50 | 4 | 0.50 | 6 | 0.50 | 8 | 0.50 | 10 | 0.50 | 12 |
| 0001 | 0.50 | 1 | 0.50 | 1 | 0.66 | 4 | 0.66 | 8 | 0.66 | 14 | 0.66 | 24 |
| 1100 | - | | 0.33 | 27 | 0.39 | 1 | 0.39 | 1 | 0.39 | 4 | 0.46 | 5 |
| 1010 | - | | 0.50 | 3 | 0.50 | 19 | 0.58 | 1 | 0.58 | 8 | 0.59 | 1 |
| 1001 | - | | 0.66 | 2 | 0.66 | 4 | 0.66 | 7 | 0.66 | 12 | 0.69 | 1 |
| 0110 | - | | 0.33 | 1 | 0.33 | 10 | 0.33 | 81 | 0.38 | 1 | 0.46 | 1 |
| 0101 | - | | 0.29 | 1 | 0.38 | 1 | 0.46 | 1 | 0.46 | 4 | 046 | 1 |
| 0011 | - | | 0.25 | 4 | 0.25 | 32 | 0.50 | 2 | 0.50 | 18 | 0.50 | 133 |
| 1110 | - | | - | | 0.33 | 2 | 0.33 | 24 | 0.33 | 221 | 0.38 | 6 |
| 1101 | - | | - | | 0.29 | 2 | 0.38 | 2 | 0.46 | 2 | 0.46 | 14 |
| 1011 | - | | - | | 0.33 | 3 | 0.33 | 8 | 0.39 | 1 | 0.52 | 1 |
| 0111 | - | | - | | 0.15 | 1 | 0.20 | 8 | 0.33 | 4 | 0.38 | 6 |
| 1111 | - | | - | | - | - | 0.33 | 4 | 0.38 | 4 | 0.46 | 4 |

**Table 8. CPM Additions of Substitutions And Associated β-lactam Antibiotics From Optimal Six Step Treatment Plans (*Maximum Probability for Path)**

| Mutations | Drugs associated with substitutions in optimal paths (probability) |
|---|---|
| 0000-1000 | CTT(0.38*) |
| 0000-0100 | |
| 0000-0010 | |
| 0000-0001 | |
| 1000-1100 | |
| 1000-1010 | |
| 1000-1001 | |
| 0100-1100 | SAM(1.0*) |
| 0100-0110 | CTX(1.0*), CPD(1.0*) |
| 0100-0101 | |
| 0010-1010 | CTT(0.22) |
| 0010-0110 | |
| 0010-0011 | |
| 0001-1001 | AM(1.0*), CTT(0.47), SAM(1.0*) |
| 0001-0101 | |
| 0001-0011 | |
| 1100-1110 | CAZ(0.85*), SAM(0.046), FEP(0.32) |
| 1100-1101 | AMP(1.0*),CAZ(0.15), SAM(0.95), FEP(0.68) |
| 1010-1110 | CEC(1.0*), CTT(0.47) |
| 1010-1011 | |
| 1001-1101 | |
| 1001-1011 | CTX(0.50*) |
| 0110-1110 | FEP(1.0*) |
| 0110-0111 | ZOX(1.0*), CXM(0.94), CPD(1.0*) |
| 0101-1101 | AMP(1.0*), FEP(1.0*) |
| 0101-0111 | CTX(0.58), ZOX(1.0*), CXM(0.59), CPD(0.85) |
| 0011-1011 | CTT(0.04) |
| 0011-0111 | ZOX(1.0*), CPD(1.0*) |
| 1110-1111 | AM(0.90), CRO(0.53), SAM(1.0*), CPD(0.39), FEP(0.72) |
| 1101-1111 | AMP(1.0*), SAM(1.0*), FEP(1.0*) |
| 1011-1111 | TZP(0.03) |
| 0111-1111 | CPD(1.0*) |

**Table 9. CPM Reversions of Substitutions And Associated β-lactam Antibiotics From Optimal Six Step Treatment Plans (*Maximum Probability for Path)**

| Reversions | Drugs associated with substitutions in optimal paths (probability) |
|---|---|
| 1111-1110 | CEC(1.0*), CAZ(0.74), CTT(0.29), CPR(1.0*), TZP(0.15) |
| 1111-1101 | AM(1.0*), AMC(1.0*), CAZ(0.26), TZP(0.85) |
| 1111-1011 | |
| 1111-0111 | ZOX(1.0*), CXM(1.0*) |
| 1110-1100 | TZP(0.49*) |
| 1110-1010 | AM(0.10), CRO(0.47*), CPD(0.28), FEP(0.28) |
| 1110-0110 | CAZ(1.0*), CPR(1.0*), CPD(0.33), TZP(0.51) |
| 1101-1100 | |
| 1101-1001 | |
| 1101-0101 | |
| 1011-1010 | TZP(0.30) |
| 1011-1001 | TZP(0.92*) |
| 1011-0011 | TZP(0.18) |
| 0111-0110 | |
| 0111-0101 | |
| 0111-0011 | |
| 1100-1000 | CTT(0.25) |
| 1100-0100 | CTX(1.0*), ZOX(1.0*), CXM(1.0*) |
| 1010-1000 | CTT(0.53*), TZP(0.49) |
| 1010-0010 | ZOX(1.0*), TZP(0.43) |
| 1001-1000 | CTX(0.42), CTT(0.56) |
| 1001-0001 | |
| 0110-0100 | CXM(0.58), TZP(1.0*) |
| 0110-0010 | |
| 0101-0100 | CTX(0.42), CXM(0.41), CPD(0.15) |
| 0101-0001 | |
| 0011-0010 | CTT(0.33), TZP(0.45) |
| 0011-0001 | CTT(0.20), TZP(0.55) |
| 1000-0000 | CPR(1.0*) |
| 0100-0000 | AM(0.62*) |
| 0010-0000 | TZP(0.71*) |
| 0001-0000 | CTT(0.092), CPR(0.14) |

**Table 10. EPM Additions of Substitutions and Associated β-lactam Antibiotics From Optimal Six Step Treatment Plans**

| Mutations | β-lactams associated with substitutions in optimal paths (probability) |
|---|---|
| 0000-1000 | |
| 0000-0100 | |
| 0000-0010 | |
| 0000-0001 | |
| 1000-1100 | |
| 1000-1010 | |
| 1000-1001 | |
| 0100-1100 | SAM(1.0*) |
| 0100-0110 | |
| 0100-0101 | TZP(1.0*) |
| 0010-1010 | |
| 0010-0110 | |
| 0010-0011 | |
| 0001-1001 | AM(1.0*), SAM(1.0*) |
| 0001-0101 | TZP(1.0*) |
| 0001-0011 | |
| 1100-1110 | CTT(1/4) |
| 1100-1101 | AMP(1.0*), CPR(1/4) |
| 1010-1110 | CTT(1/2) |
| 1010-1011 | |
| 1001-1101 | |
| 1001-1011 | CTX(1/2*) |
| 0110-1110 | CTT(1/3) |
| 0110-0111 | |
| 0101-1101 | AM(1/2), AMC(1/2) |
| 0101-0111 | |
| 0011-1011 | AMC(1/2*) |
| 0011-0111 | |
| 1110-1111 | SAM(1.0*) |
| 1101-1111 | |
| 1011-1111 | CTT(1/3) |
| 0111-1111 | SAM(1/2), CPD(1.0*) |

**Table 11. EPM Reversions of Substitutions and Associated β-lactam Antibiotics From Optimal Six Step Treatment Plans**

| Reversions | β-lactams associated with substitutions in optimal paths (probability) |
|---|---|
| 1111-1110 | CTT(1/3) |
| 1111-1101 | AM(1.0*) , AMC(1.0*) |
| 1111-1011 | |
| 1111-0111 | |
| 1110-1100 | TZP(1/2*) |
| 1110-1010 | |
| 1110-0110 | CAZ(1.0*), CPR(1.0*), TZP(1/2) |
| 1101-1100 | |
| 1101-1001 | CPR(1/3*) |
| 1101-0101 | CAZ(1.0*), TZP(1.0*) |
| 1011-1010 | CTT(1/3*) |
| 1011-1001 | AM(1/2*), CTT(1/3) |
| 1011-0011 | |
| 0111-0110 | |
| 0111-0101 | SAM(1/2*) |
| 0111-0011 | |
| 1100-1000 | CTT(1/4), CPR(1/4), TZP(1/3*) |
| 1100-0100 | CTX(1.0*), ZOX(1.0*), CXM(1.0*) |
| 1010-1000 | CTT(1/2*), TZP(1/3) |
| 1010-0010 | |
| 1001-1000 | CEC(1/2*), CTX(1/2*), CTT(1/2*), CPR(1/2*), TZP(1/3) |
| 1001-0001 | CEC(1/2*), CPR(1/2*) |
| 0110-0100 | TZP(1.0*) |
| 0110-0010 | |
| 0101-0100 | CEC(1/2*), AMC(1/2*) |
| 0101-0001 | AM(1/2*), CEC(1/2*) |
| 0011-0010 | |
| 0011-0001 | AMC(1/2*) |
| 1000-0000 | CPR(1.0*) |
| 0100-0000 | FEP(1/4) |
| 0010-0000 | SAM(1/2*), TZP(1/2*) |
| 0001-0000 | CEC(1/2*), CPR(1/3), FEP(1/3) |

**Table 12: Cyclical Treatment Paths showing Substitutions and Associated β-lactam Antibiotics**

| **Path length and probability (prob)** | **0000-0010/ 0010-0000** | **0000-0100/ 0100-0000** | **0100-0110/ 0110-0100** | **0100-1100/ 1100-0100** |
|---|---|---|---|---|
| **2-step (0.70)** | | | | |
| Cycle 1 | AM/TZP | | | |
| **4-step (0.62)** | | | | |
| Cycle 2 | | CEC/AM | CTX/TZP | |
| Cycle 3 | | CEC/AM | | SAM/CTX |
| Cycle 4 | | CEC/AM | | SAM/ZOX |
| Cycle 5 | | CEC/AM | | SAM/CXM |
| Cycle 6 | | CEC/AM | CPD/TZP | |
| **6-step (0.62)** | | | | |
| Cycle 7 | | CEC/AM | CTX/TZP(2x) | |
| Cycle 8 | | CEC/AM | CTX/TZP | SAM/CTX |
| Cycle 9 | | CEC/AM | CTX/TZP | SAM/ZOX |
| Cycle 10 | | CEC/AM | CTX/TZP | SAM/CXM |
| Cycle 11 | | CEC/AM | CTX/TZP, CPD/TZP | |

**Figure 1 AMP: Ampicillin 256 μg/ml**

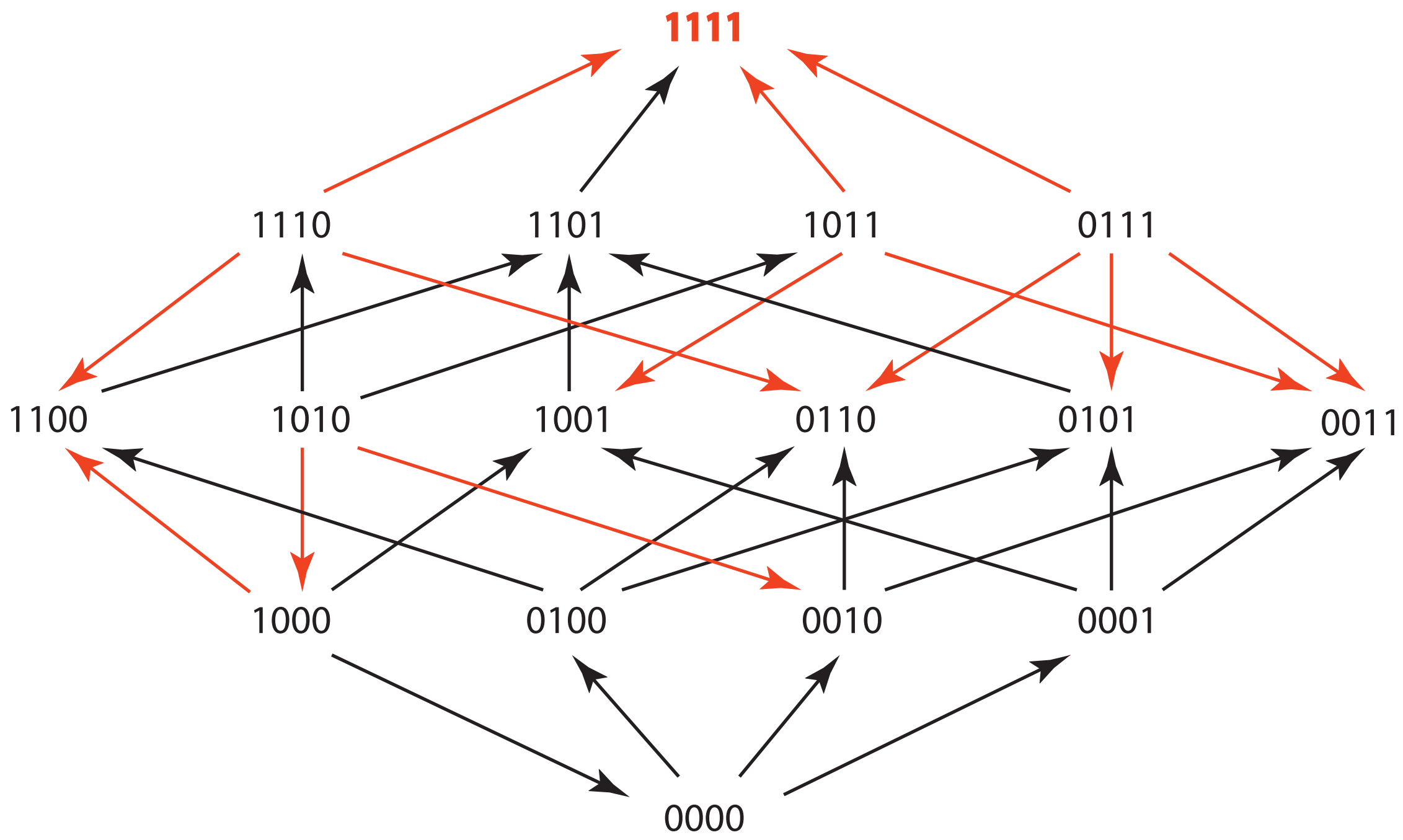

**Figure 2 AM: Amoxicillin 512 μg/ml**

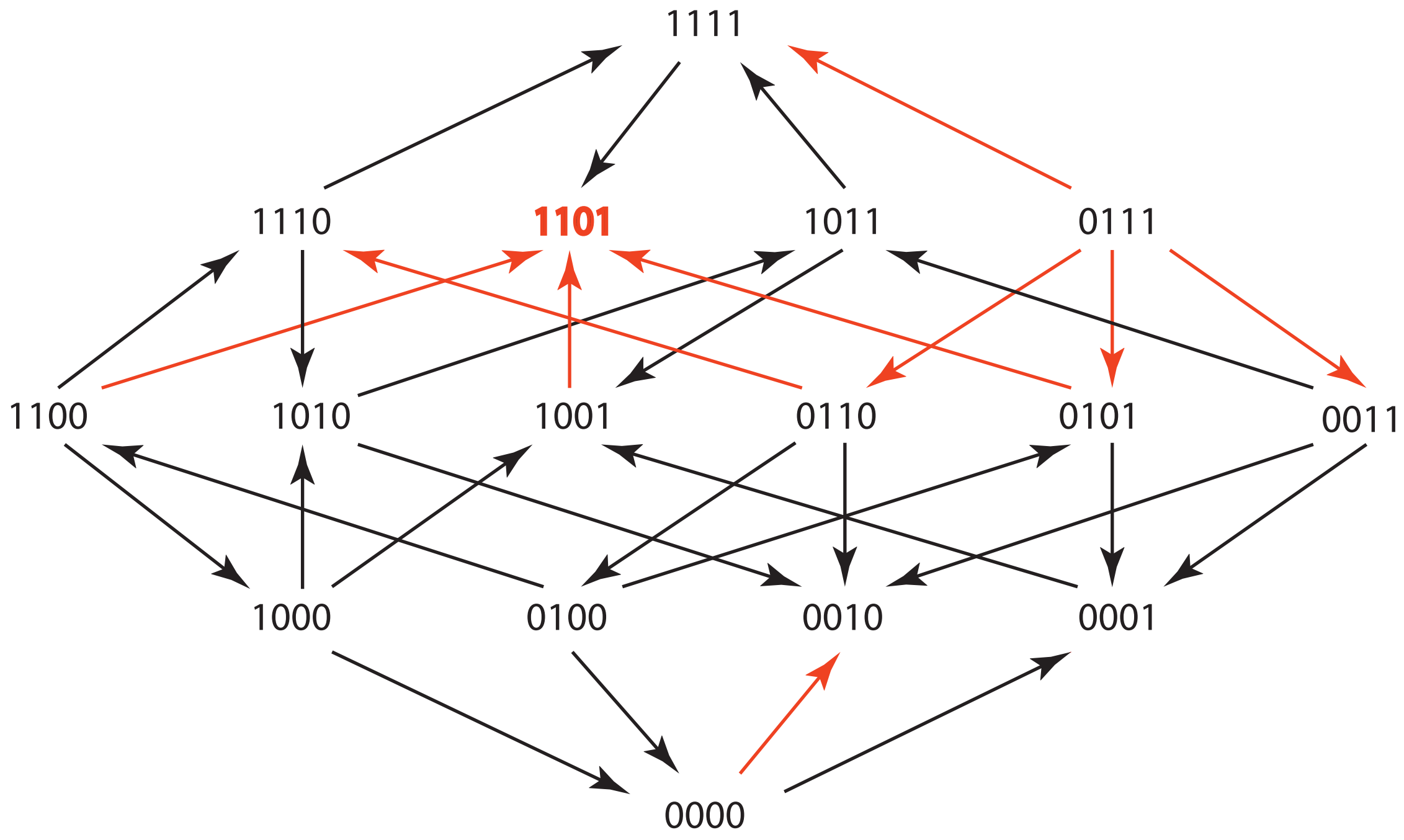

**Figure 3 CEC: Cefaclor 1 $\mu$g/ml**

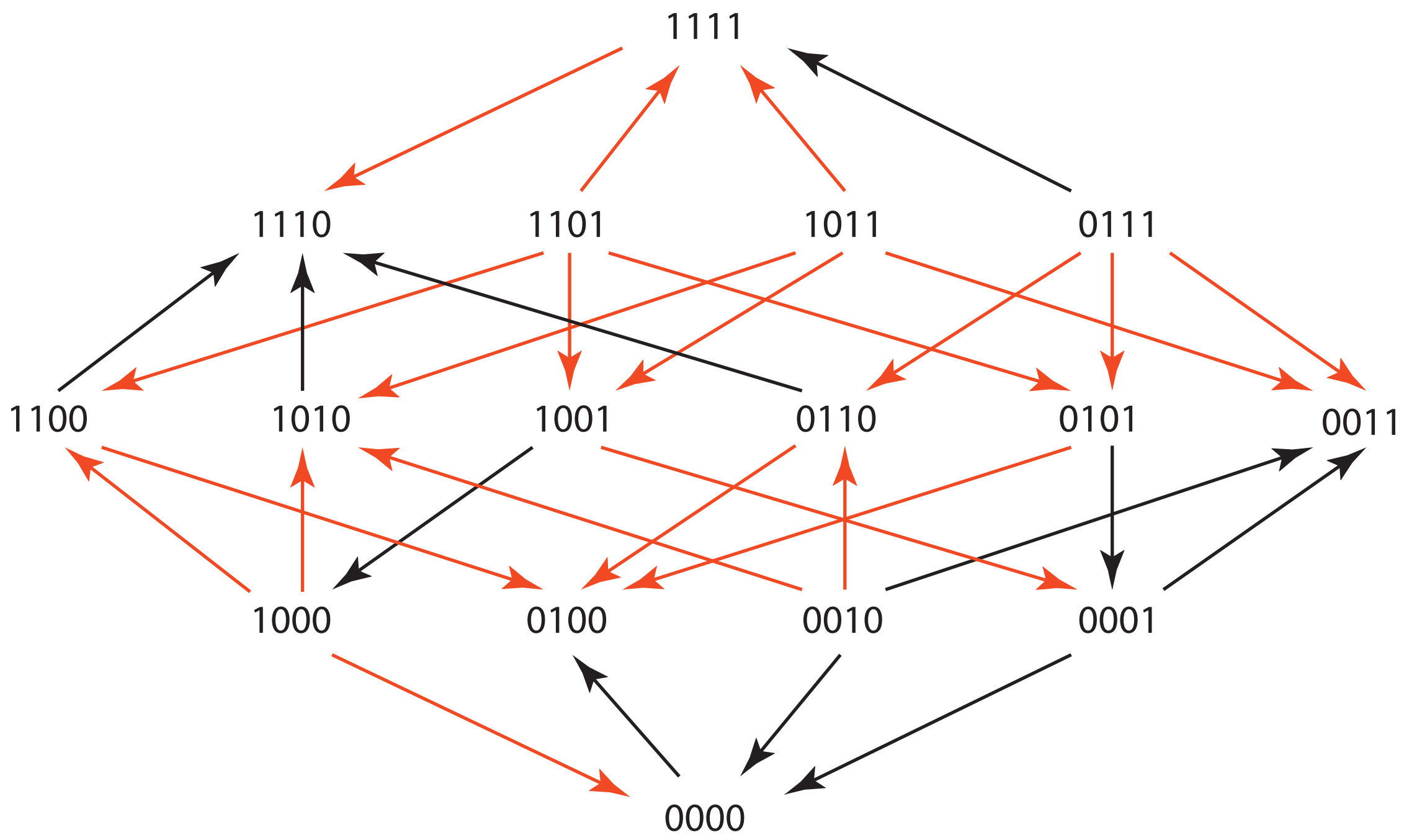

**Figure 4 CTX: Cefotaxime 0.05 $\mu$g/ml**

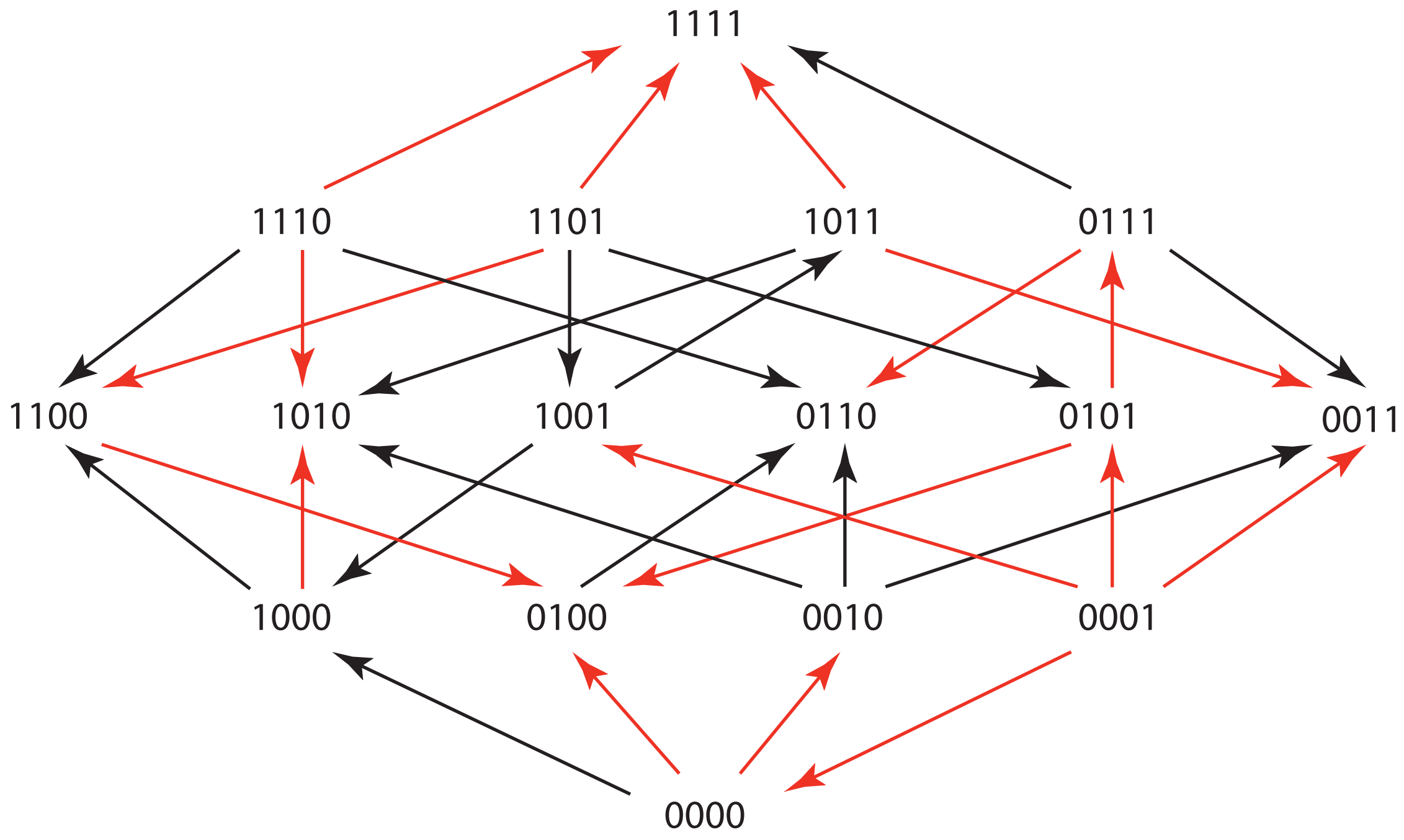

**Figure 5 ZOX: Ceftizoxime 0.03 μg/ml**

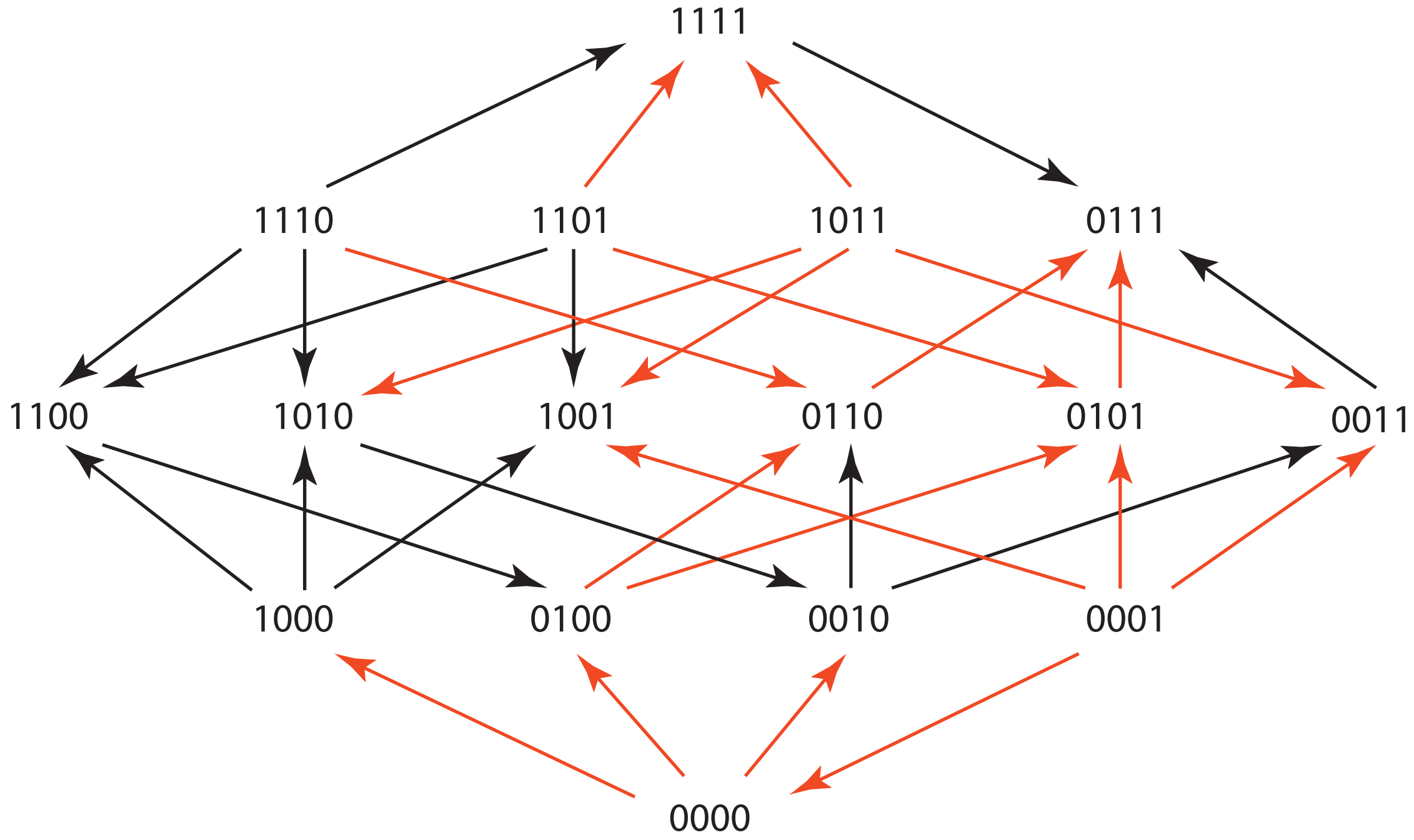

**Figure 6 CXM: Cefuroxime 1.5 μg/ml**

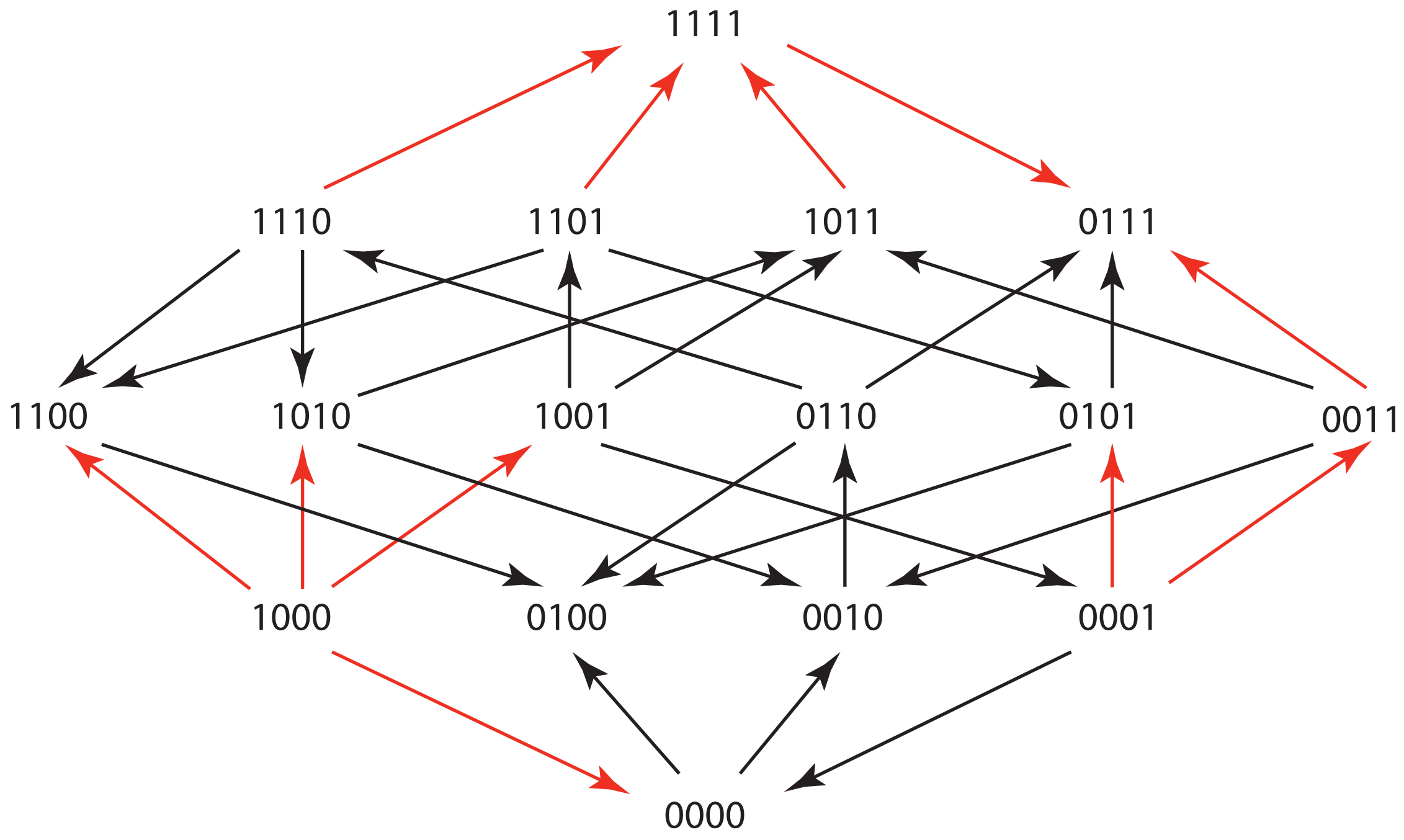

**Figure 7 CRO: Ceftriaxone 0.045 $\mu$g/ml**

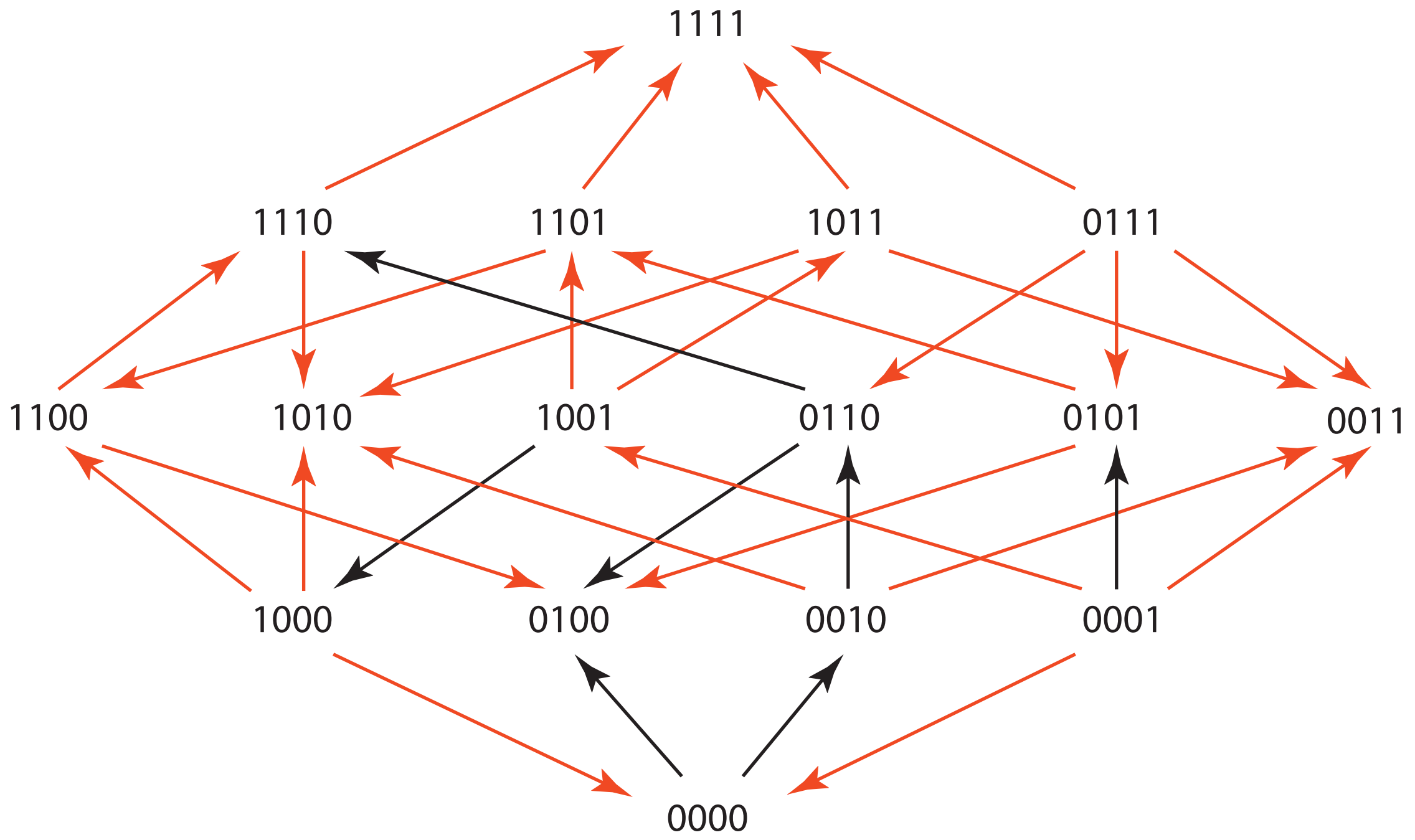

**Figure 8 AMC: Amoxicillin/Clavulanate 512 $\mu$g/ml and 8$\mu$g/ml**

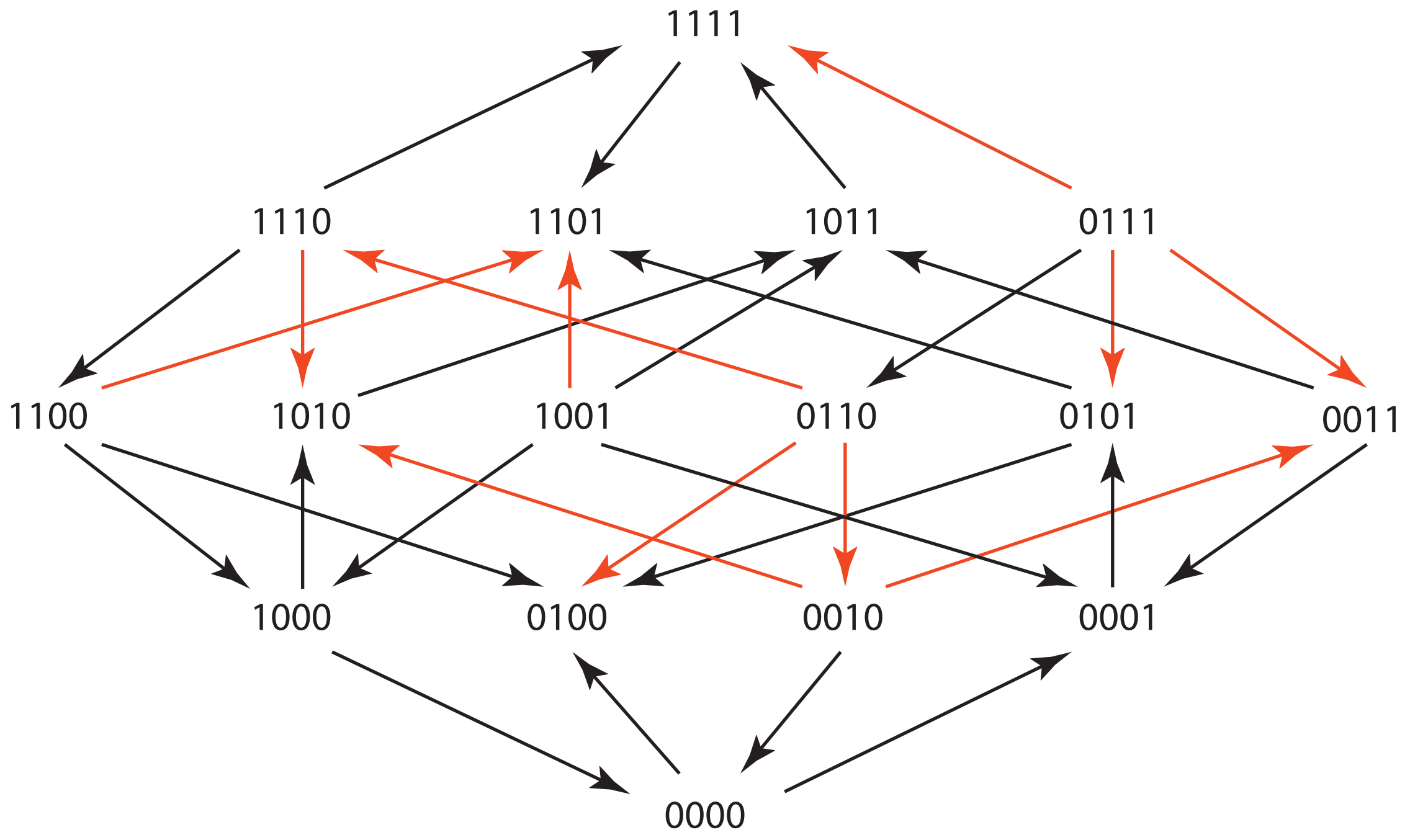

**Figure 9 CAZ: Cefazidime 0.1 $\mu$g/ml**

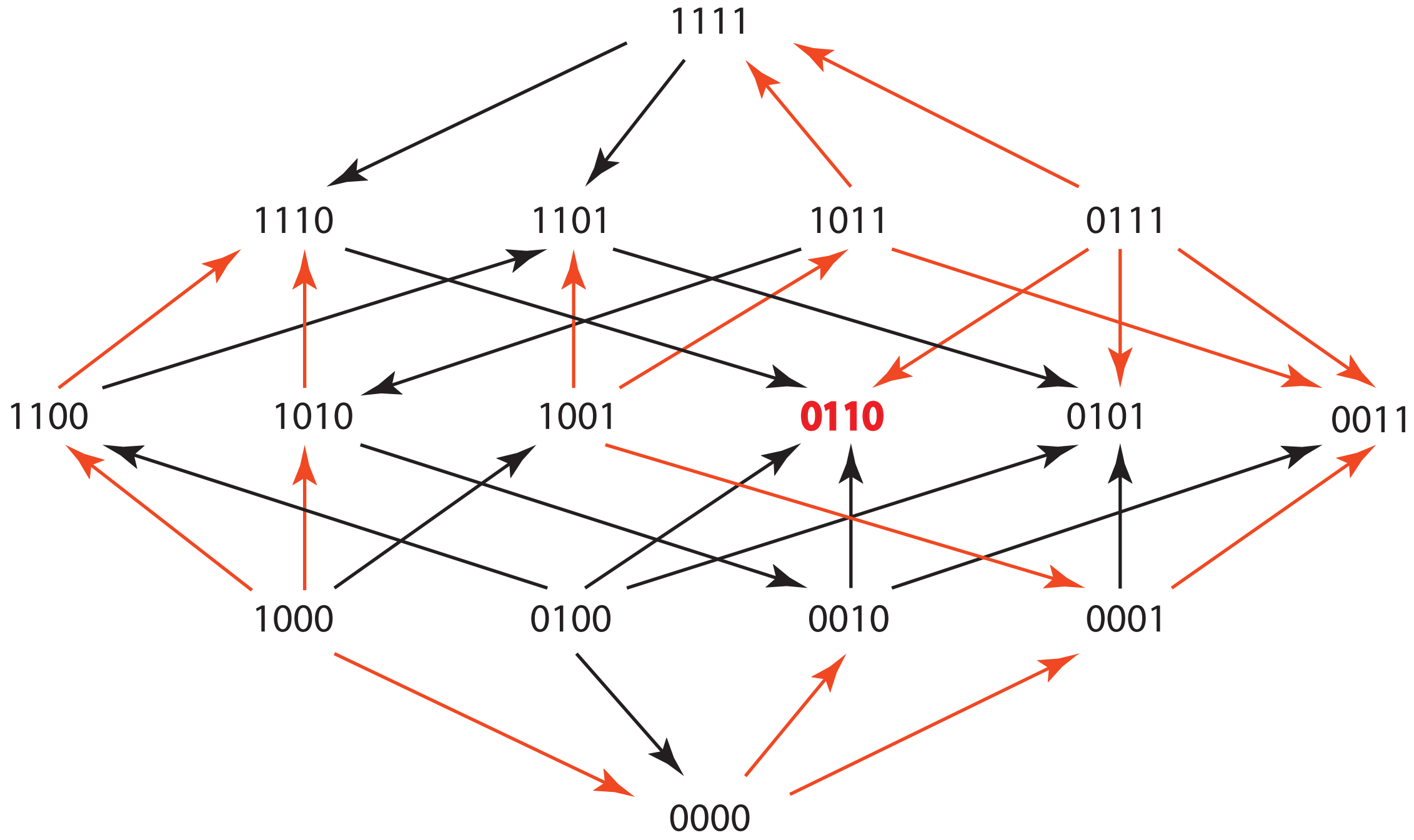

**Figure 10 CTT: Cefotetan 0.312 $\mu$g/ml**

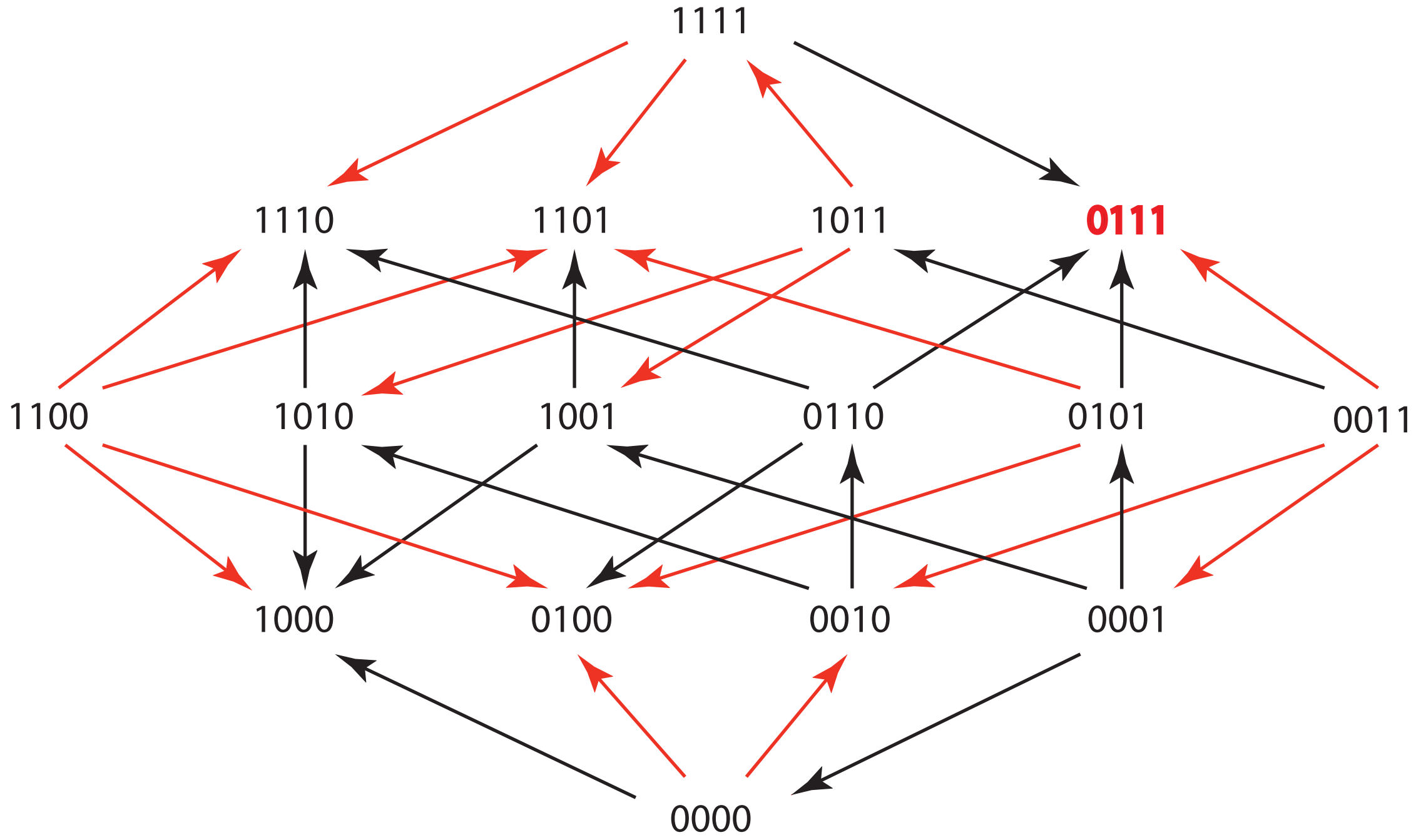

**Figure 11 SAM: Ampicillin/Sulbactam 8 $\mu$g/ml and 8$\mu$g/ml**

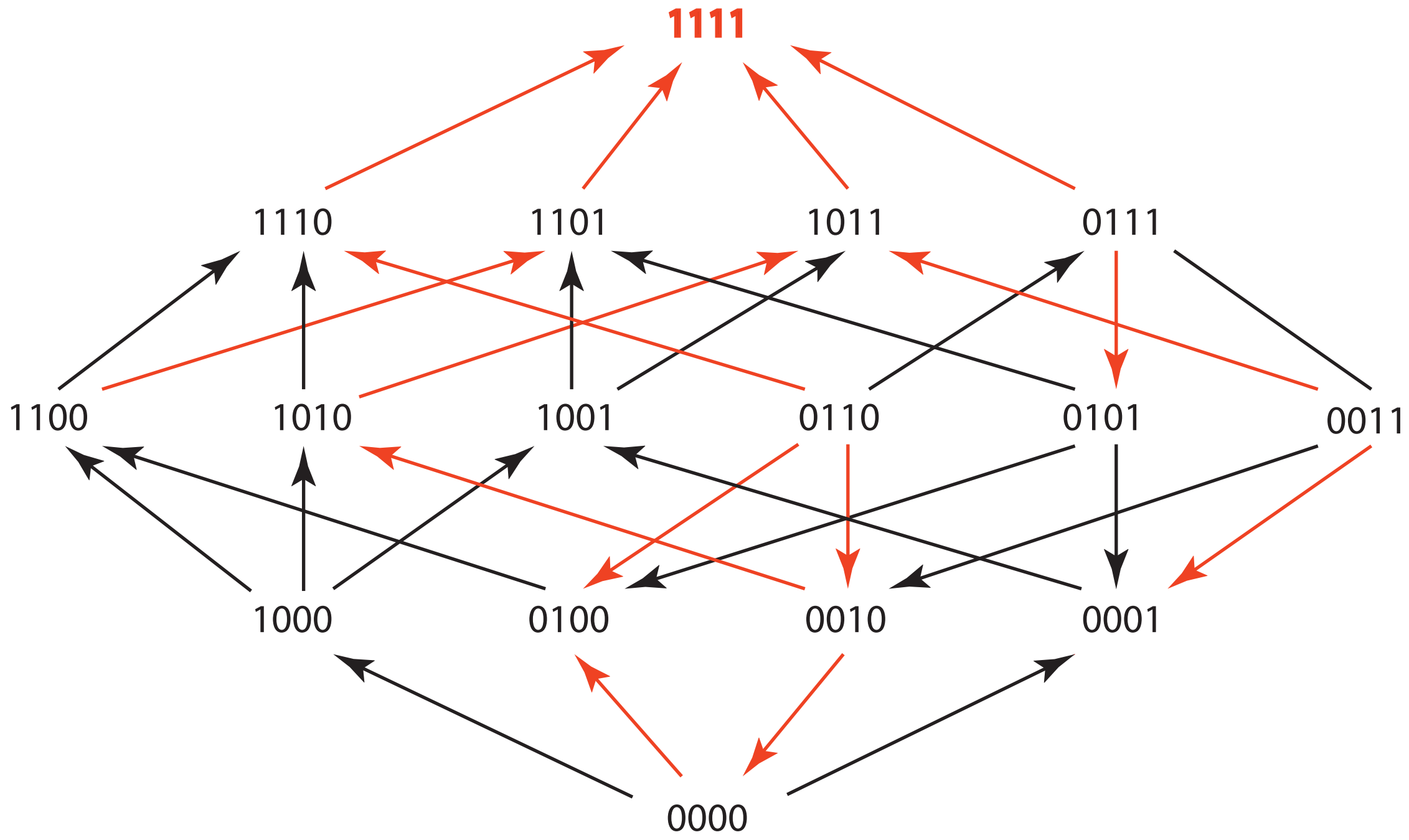

**Figure 12 CPR: Cefprozil 100 $\mu$g/ml**

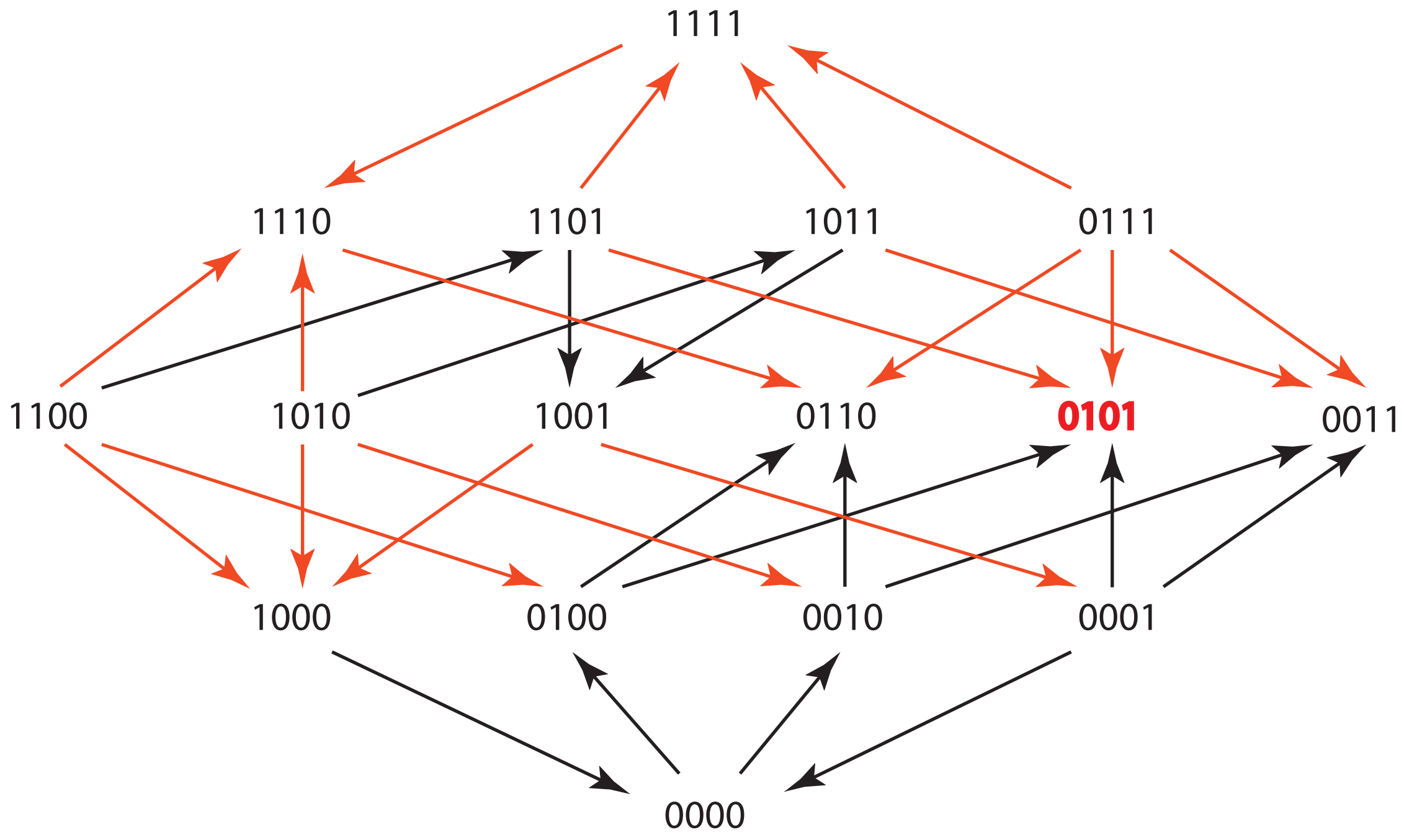

**Figure 13 CPD: Cefpodoxime 2 $\mu$g/ml**

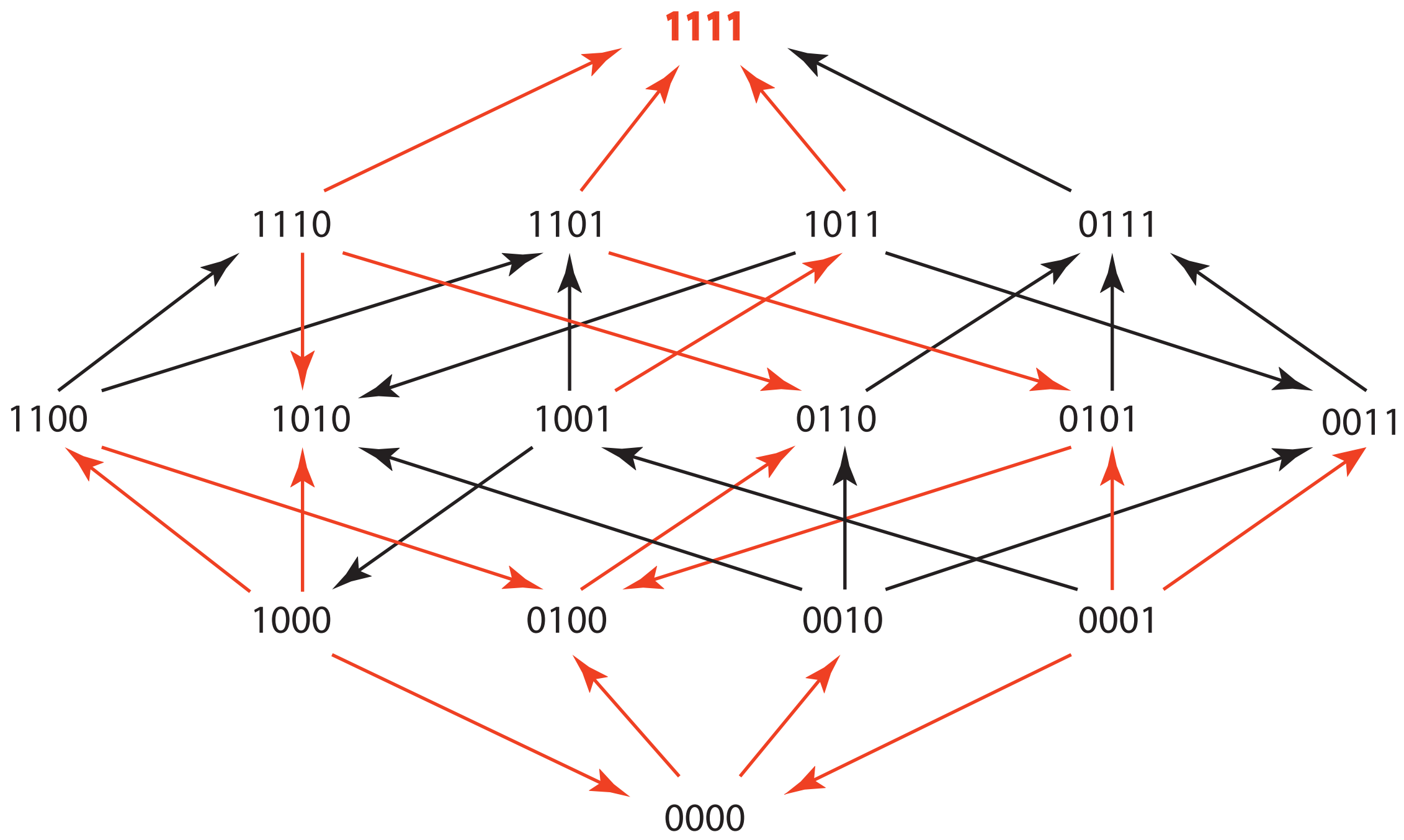

**Figure 14 TZP: Pipercillin / Tazobactam 8.12$\mu$g/ml and 8 $\mu$g.ml**

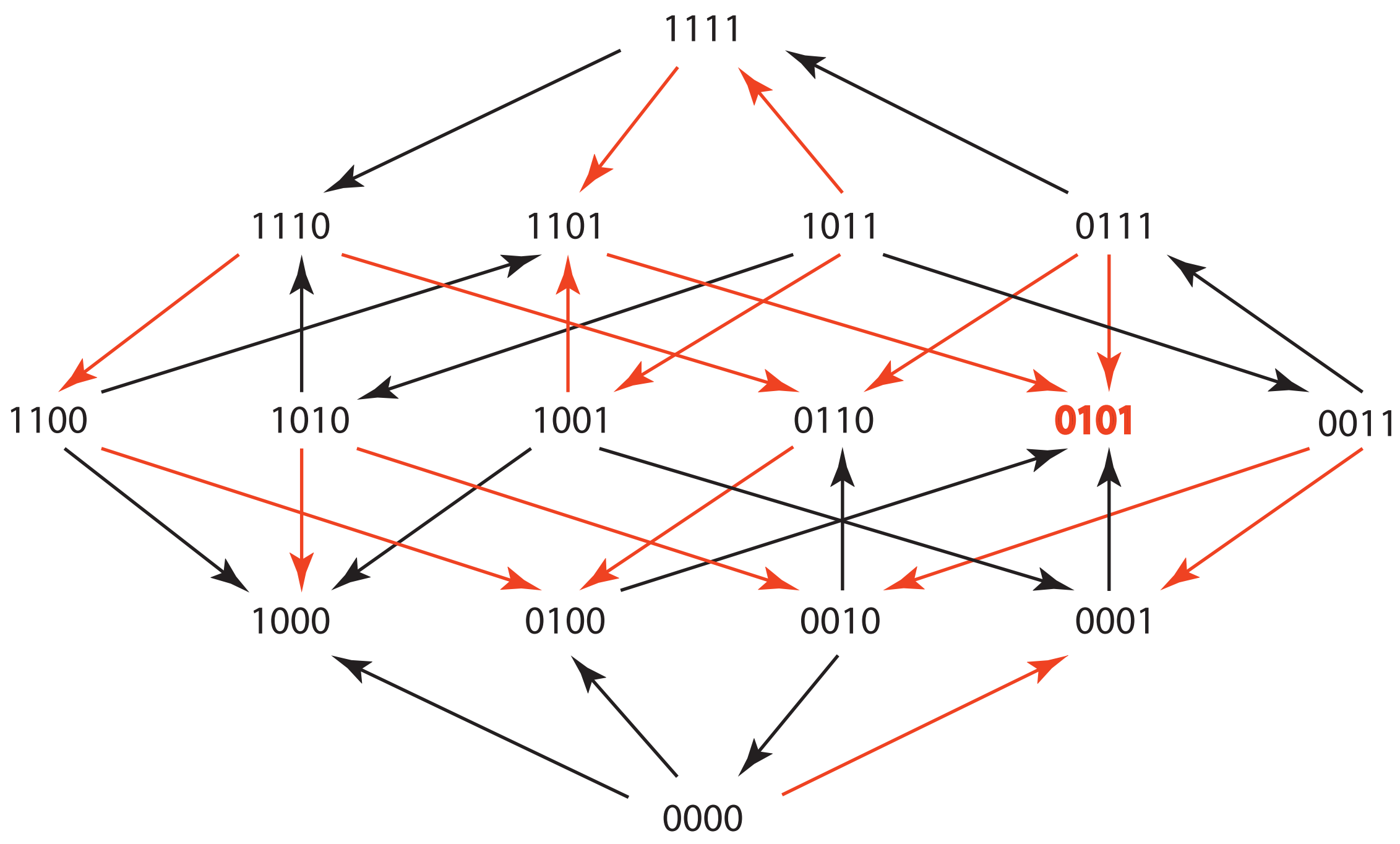

**Figure 15 FEP: Cefepime 0.0156$\mu$g/ml**

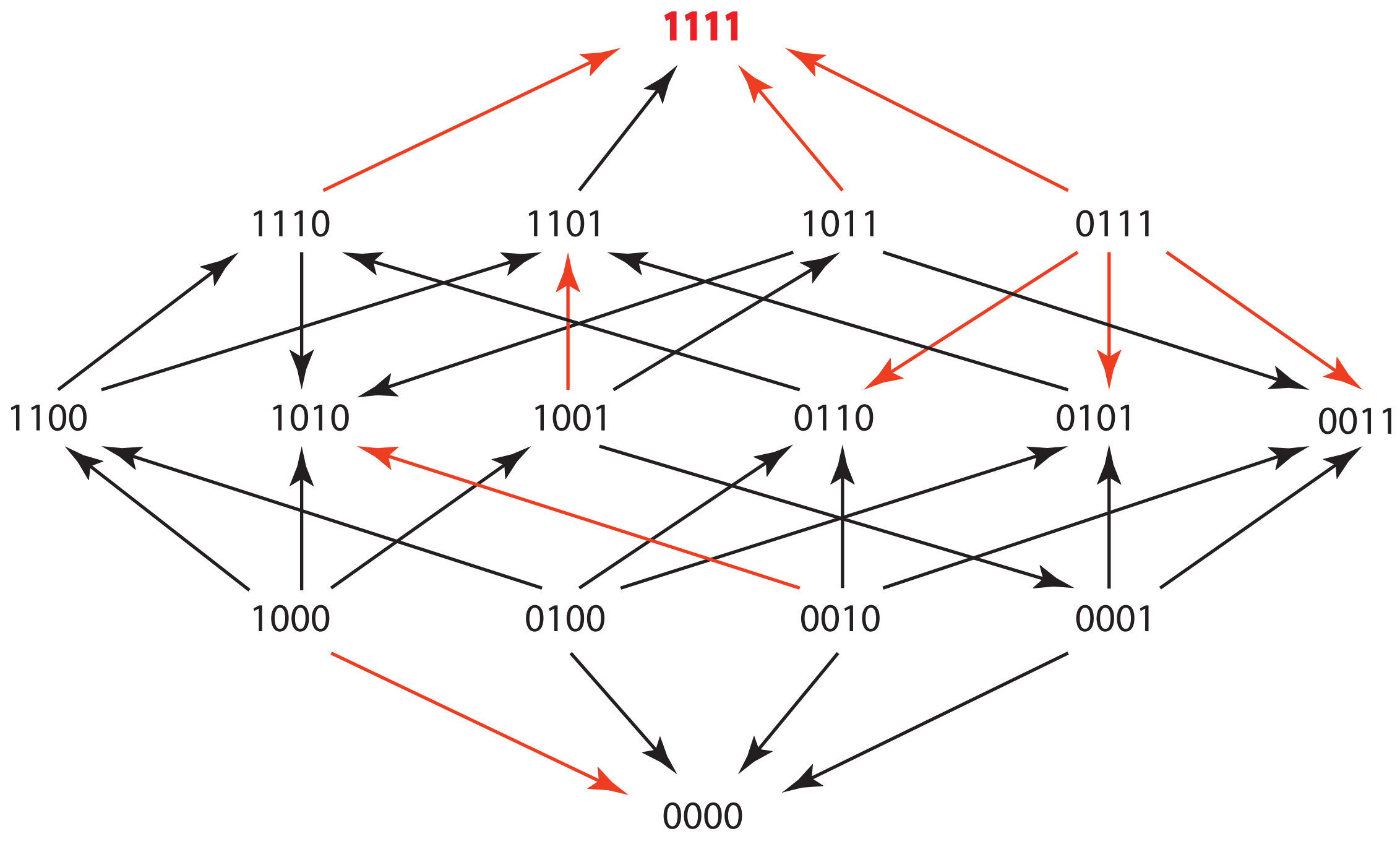

**Figure 16: Summary of Highest CPM probabilities**

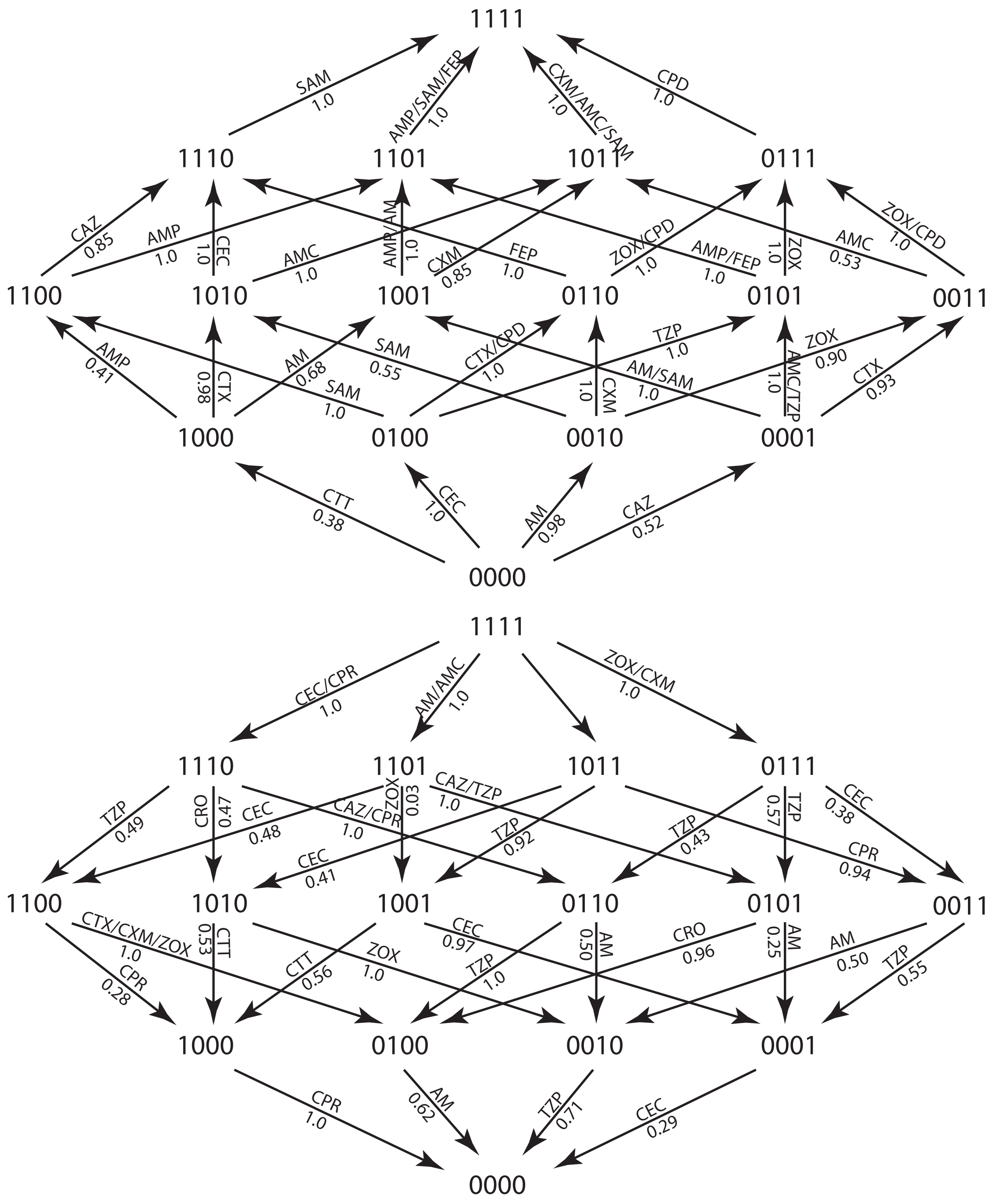

**Figure 17. Summary of Optimal Six Step Sequences (EPM and CPM)**

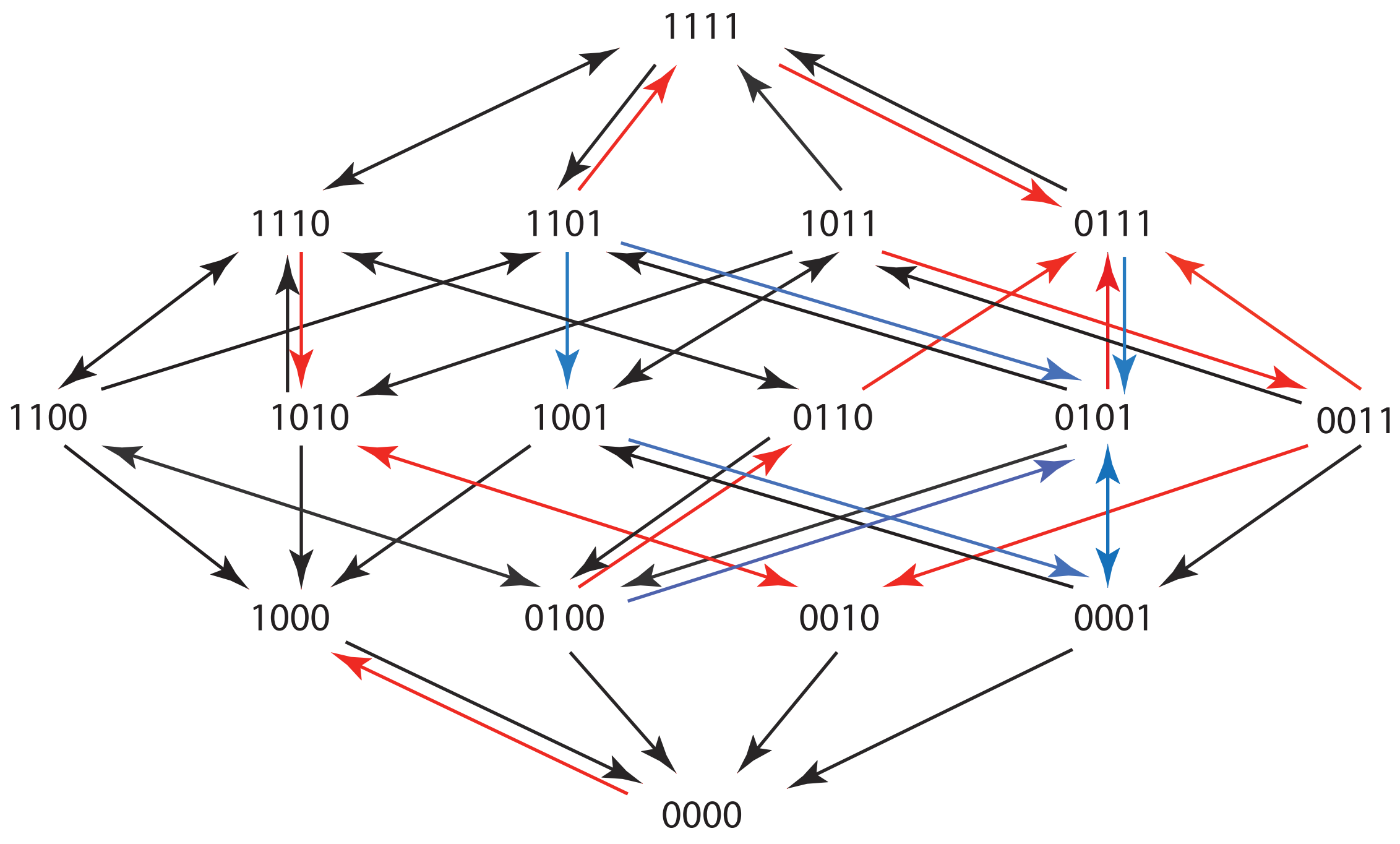

**Figure 18. Summary of Optimized Two, Four, and Six Step CPM Antibiotic Cycles**

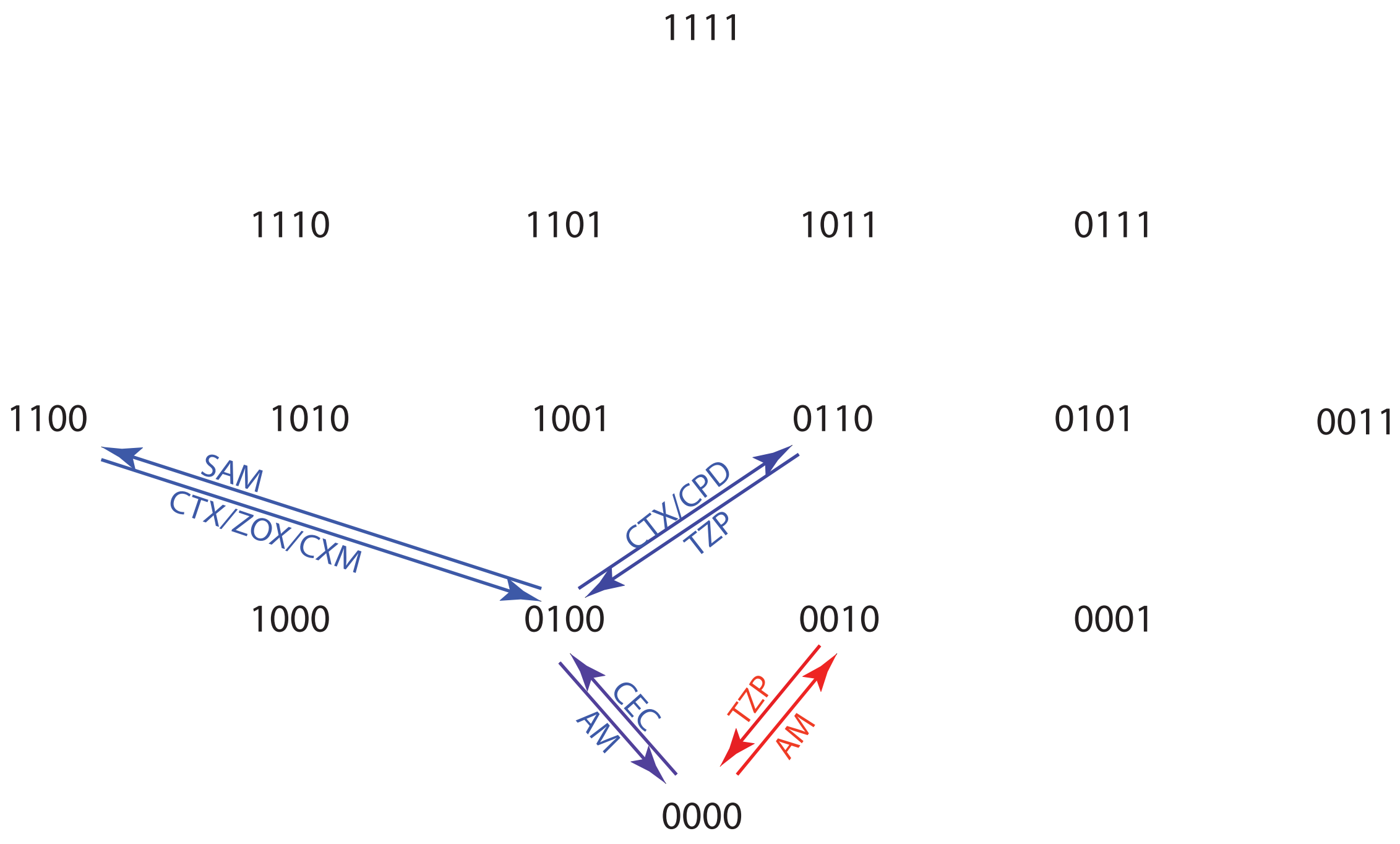

Supporting Information

For any biallelic system and set of drugs, the maximum probabilities for returning to the wild-type depend on how many steps one allows in the treatment plan. The following example demonstrates that the maximum probabilities may increase by the number of steps indefinitely.

Consider a three-loci system where the genotypes are ordered as:
000, 100, 010, 001, 110, 101, 011, 111.
Asume that the starting point is the genotype 100 and that Drugs A and B (see below) are available. For the sequence A, the probability for ending at 000 is 0.9, for A-B-A 0.99, for A-B-A-B-A 0.999, and so forth.

| | | | | | | | |
|---|---|---|---|---|---|---|---|
| 1 | 0 | 0 | 0 | 0 | 0 | 0 | 0 |
| 0.9 | 0 | 0 | 0 | 0.1 | 0 | 0 | 0 |
| 1/3 | 0 | 0 | 0 | 1/3 | 0 | 1/3 | 0 |
| 1/3 | 0 | 0 | 0 | 0 | 1/3 | 1/3 | 0 |
| 0 | 0 | 0 | 0 | 0 | 0 | 0 | 1 |
| 0 | 1/2 | 0 | 0 | 0 | 0 | 0 | 1/2 |
| 0 | 0 | 0 | 0 | 0 | 0 | 0 | 1 |
| 0 | 0 | 0 | 0 | 0 | 0 | 0 | 1 |
| 1 | 0 | 0 | 0 | 0 | 0 | 0 | 0 |
| 1/2 | 0 | 0 | 0 | 0 | 1/2 | 0 | 0 |
| 1/3 | 0 | 0 | 0 | 1/3 | 0 | 1/3 | 0 |
| 1/3 | 0 | 0 | 0 | 0 | 1/3 | 1/3 | 0 |
| 0 | 1 | 0 | 0 | 0 | 0 | 0 | 0 |
| 0 | 0 | 0 | 0 | 0 | 1 | 0 | 0 |
| 0 | 0 | 0 | 0 | 0 | 0 | 0 | 1 |
| 0 | 0 | 0 | 0 | 1/2 | 1/2 | 0 | 0 |

```
interface(quiet=true):
with(linalg): die := rand(0..10000):

alleins := transpose(array([[1,1,1,1,1,1,1,1,1,1,1,1,1,1,1,1]])):

antibiotics := [AMP,AM,CEC,CTX,ZOX,CXM,CRO,
AMC,CAZ,CTT,SAM,CPR,CPD,TZP,FEP]:

labels2 := [
[0,0,0,0],[1,0,0,0],[0,1,0,0],[0,0,1,0],
[0,0,0,1],[1,1,0,0],[1,0,1,0],[1,0,0,1],
[0,1,1,0],[0,1,0,1],[0,0,1,1],[1,1,1,0],
[1,1,0,1],[1,0,1,1],[0,1,1,1],[1,1,1,1]]:

Data :=
[[0.001850833,0.001570000,0.002024167,0.001948333,0.002081667,0.002185
833,0.000050800,0.002165000,0.002032500,0.002197500,0.002434167,0.0000
87500,0.002321667,0.000082500,0.000034200,0.002820833],
[0.001777500,0.001720000,0.001448333,0.002041667,0.001781667,0.0015566
67,0.001799167,0.002008333,0.001184167,0.001544167,0.001751667,0.00176
7500,0.002246667,0.002005000,0.000062500,0.002046667],
[0.002258333,0.000234167,0.002395833,0.002150833,0.001995833,0.0021500
00,0.002241667,0.000171667,0.002230000,0.001845833,0.002647500,0.00264
0000,0.000095000,0.000093300,0.000214167,0.000515833],
[0.000160000,0.000185000,0.001653333,0.001935833,0.000085000,0.0002250
00,0.001969167,0.000140000,0.002295000,0.000137500,0.002347500,0.00011
9167,0.000091700,0.000203333,0.002269167,0.002411667],
[0.000993333,0.001105833,0.001697500,0.002069167,0.000805000,0.0011158
33,0.001894167,0.001170833,0.002137500,0.002010000,0.002682500,0.00110
3333,0.001105000,0.000680833,0.002688333,0.002590833],
[0.001747500,0.000422500,0.002940000,0.002070000,0.001700000,0.0020241
67,0.001910833,0.001578333,0.002918333,0.002173333,0.001937500,0.00159
0833,0.001677500,0.002754167,0.003271667,0.002923333],
[0.001091667,0.000830000,0.002880000,0.002554167,0.000286667,0.0014066
67,0.003172500,0.000540000,0.002731667,0.000655833,0.003041667,0.00274
0000,0.000750833,0.001152500,0.000435833,0.003226667],
[0.001435000,0.001416667,0.001671667,0.001060833,0.001573333,0.0013766
67,0.001537500,0.001350833,0.000073300,0.001625000,0.001456667,0.00130
6667,0.001914167,0.001590000,0.000067500,0.001727500],
[0.002134167,0.000288333,0.002041667,0.002618333,0.002655833,0.0026300
00,0.001604167,0.000575833,0.002924167,0.002755833,0.002687500,0.00289
3333,0.002676667,0.001378333,0.000250833,0.002562500],
[0.002125000,0.003238333,0.003290833,0.002804167,0.001921667,0.0005458
33,0.002882500,0.002965833,0.003081667,0.002887500,0.000587500,0.00319
2500,0.003180833,0.000890000,0.003507500,0.002543333],
[0.001879167,0.002197500,0.002455833,0.000133333,0.002532500,0.0025041
67,0.002308333,0.002570000,0.000083300,0.002436667,0.000094200,0.00252
8333,0.003001667,0.002885833,0.000094200,0.003453333],
[0.001743333,0.001553333,0.002017500,0.001762500,0.001661667,0.0002225
00,0.000165000,0.000255833,0.002041667,0.002050000,0.001785000,0.00181
```

```
0833,0.000239167,0.000220833,0.000217500,0.000288333],
[0.000595000,0.000431667,0.001760833,0.002604167,0.000245000,0.0006375
00,0.002650833,0.000388333,0.002910000,0.001470833,0.003042500,0.00096
2500,0.000985833,0.001102500,0.003095833,0.003268333],
[0.002679167,0.002709167,0.003037500,0.002426667,0.002905833,0.0024533
33,0.000171667,0.002500000,0.002527500,0.003309167,0.000140833,0.00060
9167,0.002739167,0.000093300,0.000142500,0.000170833],
[0.002590000,0.002066667,0.002440000,0.002393333,0.002571667,0.0027350
00,0.002956667,0.002445833,0.002651667,0.002807500,0.002831667,0.00279
5833,0.002863333,0.002632500,0.000610833,0.003202500]]:

lprint(labels2);

cubegraph := {}:

for i from 1 to 16 do
for j from 1 to 16 do
agree := 0:
for k from 1 to 4 do
if (labels2[i][k] = labels2[j][k]) then agree := agree+1: fi:
od:
if (agree = 3) then cubegraph := cubegraph union {{i,j}}: fi:
od:od:


for i from 1 to 15 do
F := []:
for j from 1 to 16 do F := [F[],die()]: od:
Data := [Data[],F]:
od:

T := []:

for antibiotic from 1 to 15 do
print(antibiotic);
fitness := Data[antibiotic];
#lprint(fitness);

outdeg := []: allcount := 0:
for i from 1 to 16 do
count := 0:
for j from 1 to 16 do
if ((member({i,j},cubegraph)) and (fitness[i] < fitness[j])) then
# NEW
  count := count+fitness[j]-fitness[i]:
fi:
od:
allcount := allcount+count:
outdeg := [outdeg[],count]:
od:
```

```
M := []:
for i from 1 to 16 do
M := [M[],[0,0,0,0,0,0,0,0,0,0,0,0,0,0,0,0]]:
od:

for i from 1 to 16 do
if (outdeg[i] = 0) then M[i][i] := 1:
else
 for j from 1 to 16 do
 if ((member({i,j},cubegraph)) and (fitness[i] < fitness[j])) then
# NEW
 M[i][j] := (fitness[j]-fitness[i])/outdeg[i]:
# M[i][j] := 1/outdeg[i]:
 fi:
 od:
fi:
od:

T := [T[],M]:
for k from 1 to 16 do lprint(M[k],`,`); od:
od:
quit
```

```
> ## Transition matrices for the fitness landscapes of the 15 drugs
> ## computed with "Method 2" in Kristina's write-up of March 9.
>
> interface(quiet=true):
> with(linalg): with(combinat):
>
> an := [ AMP,AM,CEC,CTX,ZOX,CXM,CRO,AMC,CAZ,CTT,SAM,CPR,CPD,TZP,FEP]:
>
> labels2 := [
> [0,0,0,0],[1,0,0,0],[0,1,0,0],[0,0,1,0],
> [0,0,0,1],[1,1,0,0],[1,0,1,0],[1,0,0,1],
> [0,1,1,0],[0,1,0,1],[0,0,1,1],[1,1,1,0],
> [1,1,0,1],[1,0,1,1],[0,1,1,1],[1,1,1,1]]:
>
> M1 := array([
> [0, 0, .3455153608, .1943516429, .4601329963, 0, 0, 0, 0, 0, 0, 0,
0, 0, 0,
> 0],
> [.1882680171, 0, 0, 0, 0, .4128491231, 0, .3988828598, 0, 0, 0, 0,
0, 0, 0,
> 0],
> [0, 0, 0, 0, 0, .4708736733, 0, 0, .2427096804e-1, .5048553587, 0,
0, 0, 0,
> 0,0],
> [0, 0, 0, 0, 0, 0, 0, 0, .1476611445, 0, .8523388555, 0, 0, 0, 0,
0],
> [0, 0, 0, 0, 0, 0, 0, .1510569801, 0, .2099694380, .6389735818, 0,
0, 0, 0,
> 0],
> [0, 0, 0, 0, 0, 0, 0, 0, 0, 0, 0, 1.000000000, 0, 0, 0],
> [0, .4359087587, 0, .5444650175, 0, 0, 0, 0, 0, 0, 0, .
1053044461e-1, 0,
> .9095779128e-2, 0, 0],
> [0, 0, 0, 0, 0, 0, 0, 0, 0, 0, 0, 1.000000000, 0, 0, 0],
> [0, 0, 0, 0, 0, 0, 0, 0, 1, 0, 0, 0, 0, 0, 0],
> [0, 0, 0, 0, 0, 0, 0, 0, 0, 0, 0, 1.000000000, 0, 0, 0],
> [0, 0, 0, 0, 0, 0, 0, 0, 0, 1, 0, 0, 0, 0, 0],
> [0, 0, 0, 0, 0, .3096409060, 0, 0, .2870142929, 0, 0, 0, 0, 0, 0,
> .4033448011],
> [0, 0, 0, 0, 0, 0, 0, 0, 0, 0, 0, 0, 0, 0, 1.000000000],
> [0, 0, 0, 0, 0, 0, 0, .2903450680, 0, 0, .3278727083, 0, 0, 0, 0,
> .3817822238],
> [0, 0, 0, 0, 0, 0, 0, 0, .2137630774, .2314135342, .2567303866, 0,
0, 0, 0,
> .2980930019],
> [0, 0, 0, 0, 0, 0, 0, 0, 0, 0, 0, 0, 0, 0, 0, 1]]):
>
>
> M2 := array([
> [0, 0, 0, .9844708460, .1552915396e-1, 0, 0, 0, 0, 0, 0, 0, 0, 0, 0,
```

```
0],
> [.1352941176, 0, 0, 0, 0, 0, .1862752941, .6784305882, 0, 0, 0, 0,
0, 0, 0,
> 0],
> [.6171861963, 0, 0, 0, 0, .2031256152, 0, 0, 0, .1796881885, 0, 0,
0, 0, 0,
> 0],
> [0, 0, 0, 1, 0, 0, 0, 0, 0, 0, 0, 0, 0, 0, 0, 0],
> [0, 0, 0, 0, 0, 0, 0, 1.000000000, 0, 0, 0, 0, 0, 0, 0, 0],
> [0, .1534845128, 0, 0, 0, 0, 0, 0, 0, 0, 0, .1981204060, .
6483950812, 0, 0,
> 0],
> [0, 0, 0, .5408925955, 0, 0, 0, 0, 0, 0, 0, 0, 0, .4591074045, 0,
0],
> [0, 0, 0, 0, 0, 0, 0, 0, 0, 0, 0, 0, 1.000000000, 0, 0, 0],
> [0, 0, .1549361613, .5029328463, 0, 0, 0, 0, 0, 0, 0, .3421309925,
0, 0, 0,
> 0],
> [0, 0, 0, 0, .2526595745, 0, 0, 0, 0, 0, 0, 0, .7473404255, 0, 0,
0],
> [0, 0, 0, .5058142476, .5232561182e-1, 0, 0, 0, 0, 0, 0, 0, 0, .
4418601406,
> 0,0],
> [0, 0, 0, 0, 0, 0, .1018775295, 0, 0, 0, 0, 0, 0, 0, 0, .
8981224705],
> [0, 0, 0, 0, 0, 0, 0, 0, 0, 0, 0, 0, 1, 0, 0, 0],
> [0, 0, 0, 0, 0, 0, 0, .7406666667e-1, 0, 0, 0, 0, 0, 0, 0, .
9259333333],
> [0, 0, 0, 0, 0, 0, 0, 0, .1787042106, .2360594825, .2691184240, 0,
0, 0, 0,
> .3161178829],
> [0, 0, 0, 0, 0, 0, 0, 0, 0, 0, 0, 0, 1.000000000, 0, 0, 0]]):
>
> M3 := array([
> [0, 0, 1.000000000, 0, 0, 0, 0, 0, 0, 0, 0, 0, 0, 0, 0, 0],
> [.3403390232, 0, 0, 0, 0, .3221241399, .3375368369, 0, 0, 0, 0, 0,
0, 0, 0,
> 0],
> [0, 0, 1, 0, 0, 0, 0, 0, 0, 0, 0, 0, 0, 0, 0, 0],
> [.1388587490, 0, 0, 0, 0, 0, .1173311219, 0, .1022607496, 0, .
6415493795, 0,
> 0,0, 0, 0],
> [.2871466592, 0, 0, 0, 0, 0, 0, 0, 0, 0, .7128533408, 0, 0, 0, 0,
0],
> [0, 0, .3340880336, 0, 0, 0, 0, 0, 0, 0, .6659119664, 0, 0, 0,
0],
> [0, 0, 0, 0, 0, 0, 0, 0, 0, 0, 0, 1.000000000, 0, 0, 0, 0],
> [0, .3312722019e-1, 0, 0, .9668727798, 0, 0, 0, 0, 0, 0, 0, 0, 0, 0,
0],
> [0, 0, .2879880104, 0, 0, 0, 0, 0, 0, 0, 0, .7120119896, 0, 0, 0,
```

```
0],
> [0, 0, .7857142857, 0, .2142857143, 0, 0, 0, 0, 0, 0, 0, 0, 0, 0,
0],
> [0, 0, 0, 0, 0, 0, 0, 0, 0, 0, 1, 0, 0, 0, 0, 0],
> [0, 0, 0, 0, 0, 0, 0, 0, 0, 0, 0, 1, 0, 0, 0, 0],
> [0, 0, 0, 0, 0, .4775368302, 0, .1781572562e-1, 0, .4068551051, 0,
0, 0, 0,
> 0,.9779233910e-1],
> [0, 0, 0, 0, 0, 0, .4128722254, .1506053560e-1, 0, 0, .4908650329,
0, 0, 0,
> 0,.8120220614e-1],
> [0, 0, 0, 0, 0, 0, 0, 0, .3158376234, .2556469270, .3812508833, 0,
0, 0, 0,
> .4726456632e-1],
> [0, 0, 0, 0, 0, 0, 0, 0, 0, 0, 0, 1.000000000, 0, 0, 0, 0]]):
>
> M4 := array([
> [0, .7589174316e-2, .4533265780, .5390842477, 0, 0, 0, 0, 0, 0, 0,
0, 0, 0,
> 0,0],
> [0, 0, 0, 0, 0, .2192781692e-1, .9780721831, 0, 0, 0, 0, 0, 0, 0, 0,
0],
> [0, 0, 0, 0, 0, 0, 0, 0, 1.000000000, 0, 0, 0, 0, 0, 0, 0],
> [0, 0, 0, 0, 0, 0, .4145153749e-1, 0, .4466317983, 0, .5119166642,
0, 0, 0,
> 0,0],
> [.3067484663e-1, 0, 0, 0, 0, 0, 0, .2249488753e-1, 0, .
2147239264e-1,
> .9253578732, 0, 0, 0, 0, 0],
> [0, 0, 1.000000000, 0, 0, 0, 0, 0, 0, 0, 0, 0, 0, 0, 0, 0],
> [0, 0, 0, 0, 0, 0, 1, 0, 0, 0, 0, 0, 0, 0, 0, 0],
> [0, .4153858935, 0, 0, 0, 0, 0, 0, 0, 0, 0, 0, 0, .5846141065, 0,
0],
> [0, 0, 0, 0, 0, 0, 0, 0, 1, 0, 0, 0, 0, 0, 0, 0],
> [0, 0, .4155813571, 0, 0, 0, 0, 0, 0, 0, 0, 0, 0, 0, .5844186429,
0],
> [0, 0, 0, 0, 0, 0, 0, 0, 0, 0, 1, 0, 0, 0, 0, 0],
> [0, 0, 0, 0, 0, .1647420070e-1, .2879751239, 0, .3386950150, 0, 0,
0, 0, 0, 0,
> .3568556603],
> [0, 0, 0, 0, 0, .5232854159e-1, 0, .1896075438e-1, 0, .
1797934887e-1, 0, 0, 0,
> 0, 0, .9107313552],
> [0, 0, 0, 0, 0, 0, .2886134872, 0, 0, 0, .3504494278, 0, 0, 0, 0,
> .3609370850],
> [0, 0, 0, 0, 0, 0, 0, 0, .1047286614, 0, .3175670745, 0, 0, 0, 0,
> .5777042641],
> [0, 0, 0, 0, 0, 0, 0, 0, 0, 0, 0, 0, 0, 0, 0, 1]]):
>
> M5 := array([
```

```
> [0, .5944514692e-1, .3720827624, .5684720906, 0, 0, 0, 0, 0, 0, 0,
0, 0, 0, 0,
> 0],
> [0, 0, 0, 0, 0, .1158300264e-1, .9131274802, .7528951715e-1, 0, 0,
0, 0, 0, 0,
> 0, 0],
> [0, 0, 0, 0, 0, 0, 0, 0, .5847176080, .4152823920, 0, 0, 0, 0, 0,
0],
> [0, 0, 0, 0, 0, 0, 0, 0, .1002441078, 0, .8997558922, 0, 0, 0, 0,
0],
> [.5178726889e-1, 0, 0, 0, 0, 0, 0, .1005957105, 0, .3313474485, .
5162695722,
> 0, 0, 0, 0, 0],
> [0, 0, 1.000000000, 0, 0, 0, 0, 0, 0, 0, 0, 0, 0, 0, 0, 0],
> [0, 0, 0, 1.000000000, 0, 0, 0, 0, 0, 0, 0, 0, 0, 0, 0, 0],
> [0, 0, 0, 0, 0, 0, 0, 1, 0, 0, 0, 0, 0, 0, 0, 0],
> [0, 0, 0, 0, 0, 0, 0, 0, 0, 0, 0, 0, 0, 0, 1.000000000, 0],
> [0, 0, 0, 0, 0, 0, 0, 0, 0, 0, 0, 0, 0, 0, 1.000000000, 0],
> [0, 0, 0, 0, 0, 0, 0, 0, 0, 0, 0, 0, 0, 0, 1.000000000, 0],
> [0, 0, 0, 0, 0, .3759397366e-2, .2378447405, 0, .3110275756, 0, 0,
0, 0, 0, 0,
> .4473682865],
> [0, 0, 0, 0, 0, .4390275335e-2, 0, .2668005134e-1, 0, .3667681324,
0, 0, 0, 0,
> 0, .6021615409],
> [0, 0, 0, 0, 0, 0, .2160879401, .8726623557e-1, 0, 0, .3564855999,
0, 0, 0, 0,
> .3401602244],
> [0, 0, 0, 0, 0, 0, 0, 0, 0, 0, 0, 0, 0, 0, 1, 0],
> [0, 0, 0, 0, 0, 0, 0, 0, 0, 0, 0, 0, 0, 0, 1.000000000, 0]]):
>
> M6 := array([
> [0, 0, .7871287129, .2128712871, 0, 0, 0, 0, 0, 0, 0, 0, 0, 0, 0,
0],
> [.2378459379, 0, 0, 0, 0, .2875094263, .2671652516, .2074793841, 0,
0, 0, 0,
> 0,0, 0, 0],
> [0, 0, 1, 0, 0, 0, 0, 0, 0, 0, 0, 0, 0, 0, 0, 0],
> [0, 0, 0, 0, 0, 0, 0, 0, 1.000000000, 0, 0, 0, 0, 0, 0, 0],
> [.6263739017e-1, 0, 0, 0, 0, 0, 0, 0, 0, .6241756590, .3131869509,
0, 0, 0,
> 0,0],
> [0, 0, 1.000000000, 0, 0, 0, 0, 0, 0, 0, 0, 0, 0, 0, 0, 0],
> [0, 0, 0, .1587699164, 0, 0, 0, 0, 0, 0, 0, 0, .8412300836, 0,
0],
> [0, 0, 0, 0, .8711232734e-1, 0, 0, 0, 0, 0, 0, 0, .7100255752e-1,
> .8418851151,0, 0],
> [0, 0, .5777851259e-1, 0, 0, 0, 0, 0, 0, 0, 0, 0, 0, .9422214874,
0],
> [0, 0, .4110812809, 0, 0, 0, 0, 0, 0, 0, 0, 0, 0, .5889187191,
```

```
0],
> [0, 0, 0, .5802918014e-1, 0, 0, 0, 0, 0, 0, 0, 0, 0, .3576642751,
> .5843065447,0],
> [0, 0, 0, 0, 0, .1269532955, .9374998169e-1, 0, .3889159397, 0, 0,
0, 0, 0, 0,
> .3903807831],
> [0, 0, 0, 0, 0, .1660017823, 0, 0, 0, .2374300459, 0, 0, 0, 0, 0,
> .5965681718],
> [0, 0, 0, 0, 0, 0, 0, 0, 0, 0, 0, 0, 0, 0, 0, 1.000000000],
> [0, 0, 0, 0, 0, 0, 0, 0, 0, 0, 0, 0, 0, 0, 1, 0],
> [0, 0, 0, 0, 0, 0, 0, 0, 0, 0, 0, 0, 0, 0, 1.000000000, 0]]):
>
> M7 := array([
> [0, 0, .5501153089, .4498846911, 0, 0, 0, 0, 0, 0, 0, 0, 0, 0, 0,
0],
> [.8226364532e-1, 0, 0, 0, 0, .1812942769, .7364420778, 0, 0, 0, 0,
0, 0, 0,
> 0,0],
> [0, 0, 1, 0, 0, 0, 0, 0, 0, 0, 0, 0, 0, 0, 0, 0],
> [0, 0, 0, 0, 0, 0, .4818180472, 0, .1383117242, 0, .3798702285, 0,
0, 0, 0,
> 0],
> [.1924686653, 0, 0, 0, 0, 0, 0, .6056976941e-1, 0, .8826445625e-1,
> .6586971091,0, 0, 0, 0, 0],
> [0, 0, .5249406235, 0, 0, 0, 0, 0, 0, 0, .4750593765, 0, 0, 0,
0],
> [0, 0, 0, 0, 0, 0, 1, 0, 0, 0, 0, 0, 0, 0, 0, 0],
> [0, .2604791199, 0, 0, 0, 0, 0, 0, 0, 0, 0, 0, .1893710148, .
5501498653, 0,
> 0],
> [0, 0, .9468104120, 0, 0, 0, 0, 0, 0, 0, 0, .5318958804e-1, 0, 0, 0,
0],
> [0, 0, .9590370163, 0, 0, 0, 0, 0, 0, 0, 0, .4096298369e-1, 0, 0,
0],
> [0, 0, 0, 0, 0, 0, 0, 0, 0, 0, 1, 0, 0, 0, 0, 0],
> [0, 0, 0, 0, 0, 0, .4705347342, 0, 0, 0, 0, 0, 0, 0, 0, .
5294652658],
> [0, 0, 0, 0, 0, .2094200279, 0, 0, 0, 0, 0, 0, 0, 0, 0, .
7905799721],
> [0, 0, 0, 0, 0, 0, .3376044192, 0, 0, 0, .3157381821, 0, 0, 0, 0,
> .3466573987],
> [0, 0, 0, 0, 0, 0, 0, 0, .2901527229, .2780410040e-1, .3293312280,
0, 0, 0, 0,
> .3527119488],
> [0, 0, 0, 0, 0, 0, 0, 0, 0, 0, 0, 0, 0, 0, 0, 1]]):
>
>
> M8 := array([
> [0, 0, .6311120000, 0, .3688880000, 0, 0, 0, 0, 0, 0, 0, 0, 0, 0,
0],
```

```
> [.1317347628, 0, 0, 0, 0, 0, .8682652372, 0, 0, 0, 0, 0, 0, 0, 0,
0],
> [0, 0, 1, 0, 0, 0, 0, 0, 0, 0, 0, 0, 0, 0, 0, 0],
> [.3001336362, 0, 0, 0, 0, 0, .3823527996, 0, 0, 0, .3175135642, 0,
0, 0, 0,
> 0],
> [0, 0, 0, 0, 0, 0, 0, 0, 0, 1.000000000, 0, 0, 0, 0, 0, 0],
> [0, .4584527221e-1, .3381088825, 0, 0, 0, 0, 0, 0, 0, 0, 0, .
6160458453, 0,
> 0,0],
> [0, 0, 0, 0, 0, 0, 0, 0, 0, 0, 0, 0, 0, 1.000000000, 0, 0],
> [0, .6035193224e-1, 0, 0, .2039721864, 0, 0, 0, 0, 0, 0, 0, .
5164245738,
> .2192513075, 0, 0],
> [0, 0, .4185009846, .2585661071, 0, 0, 0, 0, 0, 0, 0, .3229329083,
0, 0, 0,
> 0],
> [0, 0, .1389585331, 0, 0, 0, 0, 0, 0, 0, 0, .8610414669, 0, 0,
0],
> [0, 0, 0, 0, .4666658667, 0, 0, 0, 0, 0, 0, 0, 0, .5333341333, 0,
0],
> [0, 0, 0, 0, 0, .9699778014e-1, .3198612655, 0, 0, 0, 0, 0, 0, 0, 0,
> .5831409544],
> [0, 0, 0, 0, 0, 0, 0, 0, 0, 0, 0, 0, 1, 0, 0, 0],
> [0, 0, 0, 0, 0, 0, 0, 0, 0, 0, 0, 0, 0, 0, 0, 1.000000000],
> [0, 0, 0, 0, 0, 0, 0, 0, .1257461571e-2, .3376717926, .3011765721,
0, 0, 0,
> 0,.3598941738],
> [0, 0, 0, 0, 0, 0, 0, 0, 0, 0, 0, 0, 1.000000000, 0, 0, 0]]):
>
>
> M9 := array([
> [0, 0, 0, .4813587160, .5186412840, 0, 0, 0, 0, 0, 0, 0, 0, 0, 0,
0],
> [.3187509228, 0, 0, 0, 0, .4043746714, .2272269889, .4964741700e-1,
0, 0, 0,
> 0,0, 0, 0, 0],
> [.4061472694e-1, 0, 0, 0, 0, .2583241529, 0, 0, .3874864490, .
3135746712, 0,
> 0,0, 0, 0, 0],
> [0, 0, 0, 0, 0, 0, 0, 0, .8155551585, 0, .1844448415, 0, 0, 0, 0,
0],
> [0, 0, 0, 0, 0, 0, 0, 0, 0, .7594917481, .2405082519, 0, 0, 0, 0,
0],
> [0, 0, 0, 0, 0, 0, 0, 0, 0, 0, 0, .8494612903, .1505387097, 0, 0,
0],
> [0, 0, 0, .4403038728, 0, 0, 0, 0, 0, 0, 0, .5596961272, 0, 0, 0,
0],
> [0, 0, 0, 0, .4173912485, 0, 0, 0, 0, 0, 0, 0, .4215719837, .
1610367678, 0,
```

```
> 0],
> [0, 0, 0, 0, 0, 0, 0, 0, 1, 0, 0, 0, 0, 0, 0, 0],
> [0, 0, 0, 0, 0, 0, 0, 0, 0, 1, 0, 0, 0, 0, 0, 0],
> [0, 0, 0, 0, 0, 0, 0, 0, 0, 0, 1, 0, 0, 0, 0, 0],
> [0, 0, 0, 0, 0, 0, 0, 0, 1.000000000, 0, 0, 0, 0, 0, 0, 0],
> [0, 0, 0, 0, 0, 0, 0, 0, 0, 1.000000000, 0, 0, 0, 0, 0, 0],
> [0, 0, 0, 0, 0, 0, .8305261021e-1, 0, 0, 0, .4814586668, 0, 0, 0, 0,
> .4354887230],
> [0, 0, 0, 0, 0, 0, 0, 0, .2693082916, .2523505370, .2454667568, 0,
0, 0, 0,
> .2328744147],
> [0, 0, 0, 0, 0, 0, 0, 0, 0, 0, 0, .7434449438, .2565550562, 0, 0,
0]]):
>
>
> M10 := array([
> [0, .3763379579, .3940844388, .2295776033, 0, 0, 0, 0, 0, 0, 0, 0,
0, 0, 0,
> 0],
> [0, 1, 0, 0, 0, 0, 0, 0, 0, 0, 0, 0, 0, 0, 0, 0],
> [0, 0, 1, 0, 0, 0, 0, 0, 0, 0, 0, 0, 0, 0, 0, 0],
> [0, 0, 0, 0, 0, 0, .2201397847, 0, .7798602153, 0, 0, 0, 0, 0, 0,
0],
> [.9186737462e-1, 0, 0, 0, 0, 0, 0, .4717620312, 0, .4363705942, 0,
0, 0, 0, 0,
> 0],
> [0, .2511855632, .2560833318, 0, 0, 0, 0, 0, 0, 0, 0, .2469097645,
> .2458213404, 0, 0, 0],
> [0, .5344177894, 0, 0, 0, 0, 0, 0, 0, 0, 0, .4655822106, 0, 0, 0,
0],
> [0, .5589743590, 0, 0, 0, 0, 0, 0, 0, 0, 0, 0, .4410256410, 0, 0,
0],
> [0, 0, .2804465349, 0, 0, 0, 0, 0, 0, 0, 0, .1486031707, 0, 0, .
5709502944,
> 0],
> [0, 0, .3063290159, 0, 0, 0, 0, 0, 0, 0, 0, 0, .2227846698, 0, .
4708863144,
> 0],
> [0, 0, 0, .3272637965, .1969734550, 0, 0, 0, 0, 0, 0, 0, 0, .
4466042868e-1,
> .4311023198, 0],
> [0, 0, 0, 0, 0, 0, 0, 0, 0, 0, 0, 1, 0, 0, 0, 0],
> [0, 0, 0, 0, 0, 0, 0, 0, 0, 0, 0, 0, 1, 0, 0, 0],
> [0, 0, 0, 0, 0, 0, .3482377336, .3628021978, 0, 0, 0, 0, 0, 0, 0,
> .2889600686],
> [0, 0, 0, 0, 0, 0, 0, 0, 0, 0, 0, 0, 0, 0, 1, 0],
> [0, 0, 0, 0, 0, 0, 0, 0, 0, 0, 0, .2884117620, .2832283500, 0, .
4283598879,
> 0]]):
>
```

```
> M11 := array([
> [0, .2055973783, .3724433778, 0, .4219592439, 0, 0, 0, 0, 0, 0, 0,
0, 0, 0,
> 0],
> [0, 0, 0, 0, 0, .3881860759, .1402949367, .4715189873, 0, 0, 0, 0,
0, 0, 0,
> 0],
> [0, 0, 0, 0, 0, 1.000000000, 0, 0, 0, 0, 0, 0, 0, 0, 0, 0],
> [.4452710826, 0, 0, 0, 0, 0, .5547289174, 0, 0, 0, 0, 0, 0, 0, 0,
0],
> [0, 0, 0, 0, 0, 0, 0, 1.000000000, 0, 0, 0, 0, 0, 0, 0, 0],
> [0, 0, 0, 0, 0, 0, 0, 0, 0, 0, 0, .4632465984e-1, .9536753402, 0, 0,
0],
> [0, 0, 0, 0, 0, 0, 0, 0, 0, 0, 0, .2758620690, 0, .7241379310, 0,
0],
> [0, 0, 0, 0, 0, 0, 0, 0, 0, 0, 0, 0, .5774809365, .4225190635, 0,
0],
> [0, 0, .4863243797, .1025581844e-1, 0, 0, 0, 0, 0, 0, 0, .
5011855081, 0, 0,
> .2234293786e-2, 0],
> [0, 0, .2818533557e-1, 0, .1409310896, 0, 0, 0, 0, 0, 0, 0, .
8308835748, 0,
> 0,0],
> [0, 0, 0, .7426932971e-2, .4627575362, 0, 0, 0, 0, 0, 0, 0, 0, .
5298155309, 0,
> 0],
> [0, 0, 0, 0, 0, 0, 0, 0, 0, 0, 0, 0, 0, 0, 0, 1.000000000],
> [0, 0, 0, 0, 0, 0, 0, 0, 0, 0, 0, 0, 0, 0, 0, 1.000000000],
> [0, 0, 0, 0, 0, 0, 0, 0, 0, 0, 0, 0, 0, 0, 0, 1.000000000],
> [0, 0, 0, 0, 0, 0, 0, 0, 0, .4108437982, 0, 0, 0, 0, 0, .
5891562018],
> [0, 0, 0, 0, 0, 0, 0, 0, 0, 0, 0, 0, 0, 0, 0, 1]]):
>
> M12 := array([
> [0, 0, .9346581030, .6534189695e-1, 0, 0, 0, 0, 0, 0, 0, 0, 0, 0, 0,
0],
> [1.000000000, 0, 0, 0, 0, 0, 0, 0, 0, 0, 0, 0, 0, 0, 0, 0],
> [0, 0, 0, 0, 0, 0, 0, 0, .4264739619, .5735260381, 0, 0, 0, 0, 0,
0],
> [0, 0, 0, 0, 0, 0, 0, 0, .9254144471, 0, .7458555294e-1, 0, 0, 0, 0,
0],
> [.1376396351, 0, 0, 0, 0, 0, 0, 0, 0, .6544952910, .2078650739, 0,
0, 0, 0,
> 0],
> [0, .2813105007, .3794257798, 0, 0, 0, 0, 0, 0, 0, 0, .3357406613,
> .3523058201e-2, 0, 0, 0],
> [0, .2961777699, 0, .3408000727, 0, 0, 0, 0, 0, 0, 0, .3511111149,
0,
> .1191104254e-1, 0, 0],
> [0, .4799628903, 0, 0, .5200371097, 0, 0, 0, 0, 0, 0, 0, 0, 0, 0,
```

```
0],
> [0, 0, 0, 0, 0, 0, 0, 0, 1, 0, 0, 0, 0, 0, 0],
> [0, 0, 0, 0, 0, 0, 0, 0, 0, 1, 0, 0, 0, 0, 0],
> [0, 0, 0, 0, 0, 0, 0, 0, 0, 0, 1, 0, 0, 0, 0],
> [0, 0, 0, 0, 0, 0, 0, 0, 1.000000000, 0, 0, 0, 0, 0, 0, 0],
> [0, 0, 0, 0, 0, 0, 0, .8880647319e-2, 0, .9649207504, 0, 0, 0, 0, 0,
> .2619860231e-1],
> [0, 0, 0, 0, 0, 0, 0, .2099999580e-1, 0, 0, .9385000123, 0, 0, 0, 0,
> .4049999190e-1],
> [0, 0, 0, 0, 0, 0, 0, 0, .3445074599, .3460812087, .2960339943, 0,
0, 0, 0,
> .1337733711e-1],
> [0, 0, 0, 0, 0, 0, 0, 0, 0, 0, 0, 1.000000000, 0, 0, 0, 0]]):
>
>
> M13 := array([
> [0, 0, .3671914961, .6328085039, 0, 0, 0, 0, 0, 0, 0, 0, 0, 0,
0],
> [.6310357404e-1, 0, 0, 0, 0, .7952341508e-1, .8573730109, 0, 0, 0,
0, 0, 0, 0,
> 0, 0],
> [0, 0, 0, 0, 0, 0, 0, 0, 1.000000000, 0, 0, 0, 0, 0, 0, 0],
> [0, 0, 0, 0, 0, 0, .5900874016e-1, 0, .3867230967, 0, .5542681631,
0, 0, 0, 0,
> 0],
> [.7749078635e-1, 0, 0, 0, 0, 0, 0, .3173424823e-1, 0, .2714021803,
> .6193727851, 0, 0, 0, 0, 0],
> [0, 0, .6252319574, 0, 0, 0, 0, 0, 0, 0, 0, .1808906052, .
1938774374, 0, 0,
> 0],
> [0, 0, 0, 0, 0, 0, 1, 0, 0, 0, 0, 0, 0, 0, 0, 0],
> [0, .3198078821e-1, 0, 0, 0, 0, 0, 0, 0, 0, 0, 0, .4409590842, .
5270601276, 0,
> 0],
> [0, 0, 0, 0, 0, 0, 0, 0, 0, 0, 0, 0, 0, 0, 1.000000000, 0],
> [0, 0, .1514360313, 0, 0, 0, 0, 0, 0, 0, 0, 0, 0, .8485639687,
0],
> [0, 0, 0, 0, 0, 0, 0, 0, 0, 0, 0, 0, 0, 0, 1.000000000, 0],
> [0, 0, 0, 0, 0, 0, .2841514484, 0, .3277700228, 0, 0, 0, 0, 0, 0,
> .3880785288],
> [0, 0, 0, 0, 0, 0, 0, 0, 0, .1752484192, 0, 0, 0, 0, 0, .
8247515808],
> [0, 0, 0, 0, 0, 0, .2738393248, 0, 0, 0, .3431098415, 0, 0, 0, 0,
> .3830508337],
> [0, 0, 0, 0, 0, 0, 0, 0, 0, 0, 0, 0, 0, 0, 1.000000000],
> [0, 0, 0, 0, 0, 0, 0, 0, 0, 0, 0, 0, 0, 0, 0, 1]]):
>
> M14 := array([
> [0, .4878056712e-1, .5826562320, 0, .3685632009, 0, 0, 0, 0, 0, 0,
0, 0, 0, 0,
```

```
> 0],
> [0, 1, 0, 0, 0, 0, 0, 0, 0, 0, 0, 0, 0, 0, 0, 0],
> [0, 0, 0, 0, 0, 0, 0, 0, 0, 1.000000000, 0, 0, 0, 0, 0, 0],
> [.7146233157, 0, 0, 0, 0, 0, 0, 0, .2853766843, 0, 0, 0, 0, 0, 0,
0],
> [0, 0, 0, 0, 0, 0, 0, 0, 0, 1.000000000, 0, 0, 0, 0, 0, 0],
> [0, .2272393379, .5188744354, 0, 0, 0, 0, 0, 0, 0, 0, 0, .
2538862267, 0, 0,
> 0],
> [0, .4851816444, 0, .4311663480, 0, 0, 0, 0, 0, 0, 0, .
8365200765e-1, 0, 0, 0,
> 0],
> [0, .2448783435, 0, 0, .4751213756, 0, 0, 0, 0, 0, 0, 0, .
2800002810, 0, 0,
> 0],
> [0, 0, 1.000000000, 0, 0, 0, 0, 0, 0, 0, 0, 0, 0, 0, 0, 0],
> [0, 0, 0, 0, 0, 0, 0, 0, 0, 1, 0, 0, 0, 0, 0, 0],
> [0, 0, 0, .4524163380, .5472537264, 0, 0, 0, 0, 0, 0, 0, 0, 0, .
3299356101e-3,
> 0],
> [0, 0, 0, 0, 0, .4901439176, 0, 0, .5098560824, 0, 0, 0, 0, 0, 0,
0],
> [0, 0, 0, 0, 0, 0, 0, 0, 0, 1.000000000, 0, 0, 0, 0, 0, 0],
> [0, 0, 0, 0, 0, 0, .3002414053e-1, .9220602935, 0, 0, .
1821094940e-1, 0, 0, 0,
> 0, .2970461658e-1],
> [0, 0, 0, 0, 0, 0, 0, 0, .4274193548, .5675030466, 0, 0, 0, 0, 0,
> .5077598566e-2],
> [0, 0, 0, 0, 0, 0, 0, 0, 0, 0, 0, .1457872968, .8542127032, 0, 0,
0]]):
>
> M15 := array([
> [1, 0, 0, 0, 0, 0, 0, 0, 0, 0, 0, 0, 0, 0, 0, 0],
> [.2126650661, 0, 0, 0, 0, .2715882271, .3616662982, .1540804086, 0,
0, 0, 0,
> 0, 0, 0, 0],
> [.1464604894, 0, 0, 0, 0, .2880389624, 0, 0, .2066723493, .
3588281989, 0, 0,
> 0, 0, 0, 0],
> [.1350114542, 0, 0, 0, 0, 0, .3867275270, 0, .1773457113, 0, .
3009153075, 0,
> 0, 0, 0, 0],
> [.3565579988e-1, 0, 0, 0, 0, 0, 0, 0, 0, .4586709351, .5056732651,
0, 0, 0, 0,
> 0],
> [0, 0, 0, 0, 0, 0, 0, 0, 0, 0, 0, .3215852743, .6784147257, 0, 0,
0],
> [0, 0, 0, 0, 0, 0, 1, 0, 0, 0, 0, 0, 0, 0, 0, 0],
> [0, 0, 0, 0, .1723751063, 0, 0, 0, 0, 0, 0, 0, .5719170248, .
2557078689, 0,
```

```
> 0],
> [0, 0, 0, 0, 0, 0, 0, 0, 0, 0, 0, 1.000000000, 0, 0, 0, 0],
> [0, 0, 0, 0, 0, 0, 0, 0, 0, 0, 0, 0, 1.000000000, 0, 0, 0],
> [0, 0, 0, 0, 0, 0, 0, 0, 0, 0, 1, 0, 0, 0, 0, 0],
> [0, 0, 0, 0, 0, 0, .2834074301, 0, 0, 0, 0, 0, 0, 0, 0, .
7165925699],
> [0, 0, 0, 0, 0, 0, 0, 0, 0, 0, 0, 0, 0, 0, 0, 1.000000000],
> [0, 0, 0, 0, 0, 0, .2964940265, 0, 0, 0, .1821648279, 0, 0, 0, 0,
> .5213411455],
> [0, 0, 0, 0, 0, 0, 0, 0, .2255064695, .2427255817, .2453959679, 0,
0, 0, 0,
> .2863719809],
> [0, 0, 0, 0, 0, 0, 0, 0, 0, 0, 0, 0, 0, 0, 0, 1]]):
>
>
>
> N1 := array([],1..16,1..16):
> N2 := array([],1..16,1..16):
> N3 := array([],1..16,1..16):
> N4 := array([],1..16,1..16):
> N5 := array([],1..16,1..16):
> N6 := array([],1..16,1..16):
> N7 := array([],1..16,1..16):
> N8 := array([],1..16,1..16):
> N9 := array([],1..16,1..16):
> N10 := array([],1..16,1..16):
> N11 := array([],1..16,1..16):
> N12 := array([],1..16,1..16):
> N13 := array([],1..16,1..16):
> N14 := array([],1..16,1..16):
> N15 := array([],1..16,1..16):
>
> for i from 1 to 16 do for j from 1 to 16 do
> N1[i,j] := M1[i,j]*cat(an[1],`_`,labels2[i][],`_`,labels2[j][]):
> N2[i,j] := M2[i,j]*cat(an[2],`_`,labels2[i][],`_`,labels2[j][]):
> N3[i,j] := M3[i,j]*cat(an[3],`_`,labels2[i][],`_`,labels2[j][]):
> N4[i,j] := M4[i,j]*cat(an[4],`_`,labels2[i][],`_`,labels2[j][]):
> N5[i,j] := M5[i,j]*cat(an[5],`_`,labels2[i][],`_`,labels2[j][]):
> N6[i,j] := M6[i,j]*cat(an[6],`_`,labels2[i][],`_`,labels2[j][]):
> N7[i,j] := M7[i,j]*cat(an[7],`_`,labels2[i][],`_`,labels2[j][]):
> N8[i,j] := M8[i,j]*cat(an[8],`_`,labels2[i][],`_`,labels2[j][]):
> N9[i,j] := M9[i,j]*cat(an[9],`_`,labels2[i][],`_`,labels2[j][]):
> N10[i,j] := M10[i,j]*cat(an[10],`_`,labels2[i][],`_`,labels2[j][]):
> N11[i,j] := M11[i,j]*cat(an[11],`_`,labels2[i][],`_`,labels2[j][]):
> N12[i,j] := M12[i,j]*cat(an[12],`_`,labels2[i][],`_`,labels2[j][]):
> N13[i,j] := M13[i,j]*cat(an[13],`_`,labels2[i][],`_`,labels2[j][]):
> N14[i,j] := M14[i,j]*cat(an[14],`_`,labels2[i][],`_`,labels2[j][]):
> N15[i,j] := M15[i,j]*cat(an[15],`_`,labels2[i][],`_`,labels2[j][]):
> od: od:
>
```

```
>
> IdentityMatrix := array([],1..16,1..16):
> for i from 1 to 16 do
> for j from 1 to 16 do
> if (i = j) then IdentityMatrix[i,j] := 1: else
IdentityMatrix[i,j] := 0: fi:
> od:od:
>
> choices := [1,2,3,4,5,6,7,8,9,10,11,12,13,14,15]:
> length2 := cartprod([choices,choices]):
> length3 := cartprod([choices,choices,choices]):
> length4 := cartprod([choices,choices,choices,choices]):
> length5 := cartprod([choices,choices,choices,choices,choices]):
> length6 :=
cartprod([choices,choices,choices,choices,choices,choices]):
>
> for L in [ length2,length3,length4,length5,length6 ] do
>
> winners := [[0,{}],[0,{}],[0,{}],[0,{}],[0,{}],[0,{}],[0,{}],[0,{}],
> [0,{}],[0,{}],[0,{}],[0,{}],[0,{}],[0,{}],[0,{}],[0,{}]]:
>
> while not L[finished] do seqq := L[nextvalue]():
> R := IdentityMatrix:
> for i from 1 to nops(seqq) do R := multiply(R,cat(M,seqq[i])): od:
> # lprint(seqq,time());
> for gg from 2 to 16 do
> if (R[gg,1] > winners[gg][1]) then
> winners[gg][1] := R[gg,1]:
> winners[gg][2] := {seqq}:
> fi:
> if (R[gg,1] = winners[gg][1]) then
> winners[gg][2] := winners[gg][2] union {seqq}:
> fi:
> od:
> end do:
>
> pathways := {}:
>
> for gg from 2 to 16 do
> if (winners[gg][1]=0) then winners[gg][1] := NONE: winners[gg][2] :=
{}: fi:
> print(` `);
>
> lprint(labels2[gg],evalf(winners[gg][1])=winners[gg]
[1],nops(winners[gg][2]));
> druglabels := {}:
> for seqq in winners[gg][2] do uu := []:
> for i from 1 to nops(seqq) do uu := [uu[],an[seqq[i]]]: od:
> druglabels := druglabels union {uu}:
>
```

```
> R := IdentityMatrix:
> for i from 1 to nops(seqq) do R := multiply(R,cat(N,seqq[i])): od:
> genfct := expand(R[gg,1]):
> mypathways := {}:
> if (type(genfct,monomial)) then
> mypathways := mypathways union {genfct}:
> else
> for j from 1 to nops(genfct) do
> mypathways := mypathways union {op(j,genfct)}:
> od:
> fi:
> lprint(uu,nops(mypathways));
> for mm in mypathways do lprint(mm); od:
> pathways := pathways union mypathways:
> od:
> od:
>
> print(` `); lprint(`Transitions used in optimal pathways:`);
>
> for transition in indets(pathways) do
> count := 0:
> for pp in pathways do
> if (member(transition,indets(pp))) then count := count + 1: fi:
> od:
> lprint(transition,count);
> od:
> print(time());
>
> od:
>
>

[1, 0, 0, 0], 1.000000000 = 1.000000000, 3 [CTT, CPR], 1
1.000000000*CTT_1000_1000*CPR_1000_0000 [CPR, FEP], 1
1.000000000*CPR_1000_0000*FEP_0000_0000 [TZP, CPR], 1
1.000000000*TZP_1000_1000*CPR_1000_0000

[0, 1, 0, 0], .6171861963 = .6171861963, 6 [AM, FEP], 1
.6171861963*AM_0100_0000*FEP_0000_0000 [CEC, AM], 1
.6171861963*CEC_0100_0100*AM_0100_0000 [CXM, AM], 1
.6171861963*CXM_0100_0100*AM_0100_0000 [CRO, AM], 1
.6171861963*CRO_0100_0100*AM_0100_0000 [AMC, AM], 1
.6171861963*AMC_0100_0100*AM_0100_0000 [CTT, AM], 1
.6171861963*CTT_0100_0100*AM_0100_0000

[0, 0, 1, 0], .7146233157 = .7146233157, 2 [AM, TZP], 1
.7146233157*AM_0010_0010*TZP_0010_0000 [TZP, FEP], 1
.7146233157*TZP_0010_0000*FEP_0000_0000

[0, 0, 0, 1], .2871466592 = .2871466592, 1 [CEC, FEP], 1
```